\documentclass[10pt,preprint]{aastex}
\def\re{{\rm Re}}
\def\im{{\rm Im}}
\def\gtwid{\mathrel{\raise.3ex\hbox{$>$\kern-.75em\lower1ex\hbox{$\sim$}}}}
\def\ltwid{\mathrel{\raise.3ex\hbox{$<$\kern-.75em\lower1ex\hbox{$\sim$}}}}
\def\haf{{\textstyle{{1}\over{2}}}}

\def\hwh{{\hat w^{\phantom{*}}}}
\def\hxh{{\hat x^{\phantom{*}}}}
\def\hyh{{\hat y^{\phantom{*}}}}
\def\hzh{{\hat z^{\phantom{*}}}}
\def\hrh{{\hat r^{\phantom{*}}}}
\def\hah{{\hat a^{\phantom{*}}}}
\def\hwc{{\hat w^* }}
\def\hxc{{\hat x^* }}
\def\hyc{{\hat y^* }}
\def\hzc{{\hat z^* }}
\def\hrc{{\hat r^* }}
\def\hac{{\hat a^* }}
\def\rwx{{\rho_{WX}}}
\def\rwy{{\rho_{WY}}}
\def\rwz{{\rho_{WZ}}}
\def\rxy{{\rho_{XY}}}
\def\rxz{{\rho_{XZ}}}
\def\ryz{{\rho_{YZ}}}
\def\rWX{{\rho_{WX}}}
\def\rWY{{\rho_{WY}}}
\def\rWZ{{\rho_{WZ}}}
\def\rXY{{\rho_{XY}}}
\def\rXZ{{\rho_{XZ}}}
\def\rYZ{{\rho_{YZ}}}
\def\la{{\langle}}
\def\ra{{\rangle}}
\def\alp{{\alpha}}
\def\ups{{\upsilon}}

\def\twr{{\tilde r}}
\def\twe{{\tilde a}}
\def\twp{{\tilde\rho}}
\def\twa{{\tilde \alpha}}

\def\ckr{{\breve r}}
\def\cke{{\breve a}}
\def\pa{{\phantom{^*}}}
\def\pmb#1{\setbox0=\hbox{#1}%
  \kern-.025em\copy0\kern-\wd0
  \kern.05em\copy0\kern-\wd0
  \kern-.025em\raise.0433em\box0 }

\shortauthors{GWINN}
\shorttitle{}
\begin{document}
\input epsf

\title{Correlation Statistics of Spectrally-Varying Quantized Noise}
\author{Carl R. Gwinn}
\affil{Department of Physics, University of California, Santa Barbara, California 93106}
\email{cgwinn@physics.ucsb.edu} 
\vskip 1 truein
\begin{abstract}

I calculate the noise in the measured correlation functions and spectra of 
digitized, noiselike signals.
In the spectral domain,
the signals are drawn from a Gaussian distribution with variance that depends on frequency.
Nearly all astrophysical signals have noiselike statistics of this type, many with
important spectral variations. 
Observation and analysis of such signals at millimeter and longer wavelengths
typically involves sampling in the time domain, and digitizing the sampled signal.
(Quantum-mechanical effects,
not discussed here, are important at infrared and shorter wavelengths.)
The digitized noise is then correlated to form a measured correlation function,
which is then Fourier transformed to produce a measured spectrum.
When averaged over many samples, 
the elements of the correlation function and of the spectrum, follow Gaussian distributions.
For each element, the mean of that distribution is the deterministic part of the measurement.
The standard deviation of the Gaussian is the noise.
Here I calculate that noise, as a function of the parameters of digitization.
The noise of the correlation function is related to the underlying spectrum,
by constants that depend on the digitization parameters.  
Noise affects variances
of elements of the correlation function and covariances between them.
In the spectral domain, noise also produces variances and covariances.
I show that noise is correlated between spectral channels, for digitized spectra, and calculate the
correlation.
These statistics of noise are important for understanding of signals sampled with very
high signal-to-noise ratio, or signals with rapidly-changing levels such as pulsars.

\end{abstract}


\section{INTRODUCTION}

\subsection{Correlation Functions and Spectra}

Electric fields from nearly all astrophysical sources are indistinguishable from
Gaussian noise.
Thus, nearly all of the information in such signals lies in variances of and covariances between
electric fields of different
polarizations, spatial locations, or frequencies.
Spectra and cross-power spectra are estimates of the variance or covariance 
of the electric field as a function of frequency.
These spectra are the Fourier transforms of auto- or cross-correlation functions.
Such correlation functions are the averaged products
of pairs of elements drawn from the series, for a range of time offsets or ``lags''.
The elements are drawn from separate series for cross-correlation, and from the same for
autocorrelation.
The resulting correlation, as a function of lag, is commonly
averaged over enough realizations to provide the desired signal-to-noise ratio. The correlation
function is then Fourier
transformed to form the desired cross-power or
autocorrelation spectrum,
as a function of frequency.  In practice,
correlation and averaging can take place before or after the Fourier transform;
this makes little difference to the result from the standpoint of this paper.

Often, data are digitized before correlation.  
For bandwidths that are within the capability of digital circuitry,
processing is usually more accurate and economical
for digital signals than for analog signals.
Digitization
involves sampling, or averaging over time intervals; and quantization, or
describing the signal amplitude in each interval as one of a discrete
set of values, rather than as a continuous variable.
Quantization is an intrinsically nonlinear operation that destroys 
information, unlike the linear operations of sampling and
Fourier transform.
I find that quantization introduces effects similar to noise in the final result,
as one might perhaps expect.

Usually observers wish to minimize noise, while maintaining an invertible,
deterministic relationship between the mean correlation and the 
underlying covariance.
Optimal parameters for quantization, and errors from departures from those parameters,
are topics of classic work in radio astronomy (see, for example, \citet{coo70,hag73,kul80,dad84} and references
therein).
Calculation of the actual noise level can be important when
signal strength varies rapidly, 
and quantizer settings cannot remain optimal,
as is sometimes the case for pulsars \citep{jen98};
or when the distribution of the intensity of the signal must be measured accurately
\citep{gwi00}.
As sensitivities of radiotelescopes improve, and as demands on the observed data increase,
calculation of the noise level from quantizer parameters can be expected to become more important.

Because correlation functions and spectra are averaged over many
realizations, the Central Limit Theorem implies
that the resulting correlation function or spectrum has Gaussian
statistics.  Thus, the statistics of the spectrum are fully described
by each spectral channel's mean and 
variance, and covariances between channels.  
The mean of the spectrum is the deterministic part
of the measurement; variances and covariances are
the random part, or noise.  In principle, one
seeks to minimize the noise, while preserving the relationship
between the mean and the underlying spectrum.
In \citet{gwi04} (hereafter Paper 1), I discussed this problem for ``white'' signals,
which have zero correlation except for elements
of the two series with zero lag.
Here I consider the more general case, where signals have arbitrary
spectral character, so that covariance can depend on lag.
The effect of quantization on the statistics of such ``colored'' spectra, 
particularly their noise, is the subject of this paper.

I calculate the noise in the quantized cross- and autocorrelation functions.
The noise differs from that for correlation of continuous data in 
additional terms, some of them constant and others proportional
to the autocorrelation function,
and products of auto- and cross-correlation functions.
I present this calculation through second order in correlation.

I then 
Fourier transform these expressions to determine the mean spectrum
and its variance. 
The mean spectrum is simply the Fourier transform of the mean correlation function,
while
the noise in the spectrum is a double Fourier transform of the noise in the correlation function.
I find that in the spectral domain, a gain factor, and white noise added in quadrature, approximately 
represent the effects of quantization in a single channel.
The added white noise is commonly known as ``quantization noise''.
Indeed, the gain factor and the noise
are identical to those previous workers found for the deterministic,
mean spectrum \citep{coo70,jen98}.
However, I also show that noise is correlated across spectral channels.  This covariance 
can reduce, or increase,
the total noise in the spectrum, depending on the details of the quantization scheme and
the details of the spectrum.  
For ``white'' signals without spectral variation, 
the more general result of this paper
reduces to that found in Paper 1.  
In this case, the correlations 
or anti-correlations 
of noise between channels can represent an effect of the same order as
quantization noise, when integrated over all channels of a spectrum.

\subsection{Organization of this Paper}

I consider cross-correlation of two time series, $x$ and $y$, and autocorrelation
of $x$.
In \S\ref{continuous} I introduce these 
underlying complex time series 
$x_{\ell}$ and $y_{\ell}$ and describe their assumed statistical
properties.
I calculate the mean and variance of their correlation
function and its Fourier transform.
I show that covariance of noise in different spectral channels is zero
for correlation of the continuous (un-digitized) series.

I introduce information-destroying quantization in 
\S\ref{quantized}. 
Under the assumption that the covariances are small (except for
the zero lag of the autocorrelation function, which must be 1),
I calculate the mean and the variance
of the quantized correlation function
for quantized data, and present analytic expressions for them.
I compare the analytical results with computer simulations and find excellent agreement.

In \S\ref{section_quantspectrum} I find the statistics of the cross-power and power spectra.
The cross-power spectrum is the Fourier transform of the cross-correlation function;
and the power spectrum (sometimes called the autocorrelation spectrum) is the Fourier
transform of the autocorrelation function.
I calculate the noise in the spectra by a double Fourier transform of the noise in the correlation functions.
I show that the noise in a single spectral channel
can be approximately represented by a gain factor and white ``digitization noise''
added in quadrature with the original signal.
However, I also show that noise is correlated (or, more commonly, anti-correlated)
across spectral channels.
I present analytical results for autocorrelation functions, and autocorrelation spectra,
in \S\ref{acf_subsection} and\ \ref{acspect_subsection}.
I summarize results in \S\ref{summary}, and show that correlations of noise between channels can
represent an effect of the same order as quantization noise, when integrated over all spectral channels.

\section{CORRELATION FUNCTIONS AND SPECTRA OF CONTINUOUS SIGNALS}\label{continuous}

\subsection{Time Series of Gaussian Noise}

Consider time series $x_{\ell}$ and $y_{\ell}$.
These might be, for example, the electric fields recorded as analog signals 
at two antennas.
All elements of each are
drawn from Gaussian
distributions in the complex plane.
The distributions have zero mean.
I further assume that the series are stationary,
so that the properties of $x_{\ell}$ and $y_{\ell}$ are independent of
the time index $\ell$.
Thus, the variance of each series is constant,
and the covariances between elements can depend only
on their time separation and whether they belong to the same
or different series.
The ensemble-average spectra, defined in \S\ref{meanspectrum} below,
depend only on these variances and covariances.

For this paper, I assume that the series are statistically identical,
in the sense that the exchange of $x$ and $y$
leaves the statistical properties of the spectra unchanged.
Instrumental effects often violate this assumption in a mild fashion,
as by variations in complex gain between two antennas.
Sometimes the assumption is violated in a more fundamental way,
as in spatial variation of the spatial and spectral character
of scintillating sources \citep{des92, jau00, den02}.
This assumption can easily be relaxed, by defining separate autocorrelation
functions for the two series in the results below.

For convenience, I scale variances of real and imaginary parts
of the series $x_{\ell}$ and $y_{\ell}$ to 1.
(This is in accord with 
much of the literature on quantization, which assumes real series with
unit variance.)
The variances are then:
\begin{equation}
\haf \la x_{\ell}\, x_{\ell}^*\ra =\haf \la y_{\ell}\, y_{\ell}^*\ra = 1 .
\label{xxc_yyc_equiv_1}
\end{equation}
Here, the angular brackets $\la ... \ra$ indicate a statistical
average, over an ensemble of time series with identical statistics.

I assume that the time series have no particular intrinsic
overall phase,
so that the transformation  
\begin{equation}
x_{\ell}\rightarrow x_{\ell}e^{i\phi},
\quad
y_{\ell}\rightarrow y_{\ell}e^{i\phi}
\label{phase_invariance}
\end{equation}
leaves the variances and covariances unchanged.
Consequently, products of factors with the same conjugation average to zero:
\begin{equation}
\la x_{\ell} y_{m} \ra =\la x_{\ell} x_{m}\ra =\la y_{\ell} y_{m}\ra =0 ,
\label{no_intrinsic_phase}
\end{equation}
for any $\ell$ and $m$.

\subsection{Mean Correlation Function and Spectrum: Continuous Data}\label{continuous_avg}

\subsubsection{Mean Correlation Function}

The covariances between elements in the series $x$ and $y$
are given by
the statistically-averaged cross-correlation $\rho_{\tau}$,
and the statistically-averaged auto-correlation $\alpha_{\tau}$:
\begin{eqnarray}
\rho_\tau &=& \haf \la x_{\ell}\, y_{\ell+\tau}^* \ra\label{rhotau_alptau_def}\\
\alp_{\tau} &=& \haf \la x_{\ell}\, x_{\ell+\tau}^* \ra
=\haf \la y_{\ell}\, y_{\ell+\tau}^* \ra . \nonumber
\end{eqnarray}
Note the conjugation symmetry of $\alp_{\tau}$:
\begin{equation}
\alp_{\tau}=\alp_{-\tau}^* .
\end{equation}
Eq.\ \ref{xxc_yyc_equiv_1} gives
$\alp_0 = 1$.

Measurements seek to estimate the statistically-averaged correlation
functions via the finite averages:
\begin{eqnarray}
r_{\tau}&=&{{1}\over{2 N_o}} \sum_{\ell=1}^{N_o} x_{\ell}\, y_{\ell+\tau}^* \label{rtau_atau_def} \\
a_{\tau}&=&{{1}\over{2 N_o}} \sum_{\ell=1}^{N_o} x_{\ell}\, x_{\ell+\tau}^* . \nonumber
\end{eqnarray}
Here, $N_o$ is the number of elements observed in each series.

I assume that the correlation functions ``wrap,''
in the sense that:
\begin{equation}
x_{(\ell)}=x_{(\ell+N_o)},
y_{(\ell)}=y_{(\ell+N_o)}\quad {\rm for\ all\ } \ell .
\label{wrap_assumption} 
\end{equation}
Then, the sums in Eq.\ \ref{rtau_atau_def}
contain the same number of terms, for each $\tau$.
This simplifies counting arguments below.
Also, of course, $r_{\tau}=r_{\tau+N_o}$;
this simplifies discussion of the Fourier transform to spectra.
Note that in practice, many correlator do not ``wrap'' in this fashion.
They zero-pad the data so that $x_{\ell}\, y_{\ell+\tau}^*=0$,
if either $\ell$ or $\ell+\tau$ is greater than $N_o$ or less than zero.
The issue is moot if the number of lags correlated
is smaller than the span of data $N_o$,
or for ``FX'' correlators, which correlate in the frequency domain.
Otherwise, it can affect the noise,
through uneven sampling of 
$\alpha$ in Eq.\ \ref{continuous_rtruc} below.
I will discuss the effect heuristically in a separate paper,
in comparison of theory with measurements.

With the definitions in Eq.\ \ref{rtau_atau_def},
\begin{eqnarray}
\la r_{\tau}\ra&=&\rho_{\tau} \label{ra_relate_rhoalp}\\
\la a_{\tau}\ra&=&\alp_{\tau} . \nonumber
\end{eqnarray}
Note that Greek letters $\rho$ and $\alpha$
denote the statistically-averaged quantities,
whereas roman letters $r$ and $a$ denote the observed,
finite averages.

\subsubsection{Mean Spectrum}\label{meanspectrum}

The statistically-averaged 
cross- and auto-correlation functions are related to the cross-power and autocorrelation spectra
by Fourier transforms:
\begin{eqnarray}
\twp_k &=& \sum_{\tau=-N}^{N-1} e^{i{{2\pi}\over{2 N}}k\tau}\rho_{\tau} \label{define_statavg_spectra}\\
\twa_k &=& \sum_{\tau=-N}^{N-1} e^{i{{2\pi}\over{2 N}}k\tau}\alp_{\tau} . \nonumber 
\end{eqnarray}
Here, $2 N$ is the number of frequency channels.
Note that $\twa_k$ is real, because of the conjugation symmetry of $\alp_{\tau}$.
Other conventions for the Fourier transform
have been used in the past.
The present convention has the advantage that
the spectrum $\twa_k$ has values that 
are independent of numbers of samples $N_o$ or of spectral channels $2N$.

Similarly, 
I define the measured
cross-power and autocorrelation spectra,
\begin{eqnarray}
\twr_k &=& \sum_{\tau=-N}^{N-1} e^{i{{2\pi}\over{2 N}}k\tau} r_{\tau} \\
\twe_k &=& \sum_{\tau=-N}^{N-1} e^{i{{2\pi}\over{2 N}}k\tau} a_{\tau} . \nonumber 
\end{eqnarray}
So, by Eqs.\ \ref{ra_relate_rhoalp} and \ref{define_statavg_spectra},
\begin{eqnarray}
\la \twr_k\ra &=& \twp_k \label{twrk_twpk_twak_twak}\\
\la \twe_k\ra &=& \twa_k . \nonumber
\end{eqnarray}

As a simple example, a ``white'' spectrum
with a spectrally-uniform correlation $\rho_w$ has
$\twa_k=1$ and
$\twp_k=\rho_w$.
Then, only the zero lags 
of the statistically-averaged correlation functions
will have nonzero values:
$\alp_0=1$
and 
$\rho_0=\rho_w$.
For all other lags $\tau$, $\alp_{\tau}=\rho_{\tau}=0$.

\subsection{Noise: Continuous Data}

\subsubsection{Noise for Correlation Function}

The variance of the observed correlation function
describes the noise.
We therefore seek:
\begin{eqnarray}
\la r_{\tau} r_{\ups}^*\ra &=&
{{1}\over{(2 N_o)^2}} \sum_{\ell=1}^{N_o}\sum_{m=1}^{N_o} \la  x_{\ell} y_{\ell+\tau}^* x_m^* y_{m+\ups}\ra .
\label{rtru_sum_expand}
\end{eqnarray}
The fourth moment of elements drawn from a Gaussian distribution 
is related to their second moments, 
so that:
\begin{equation}
\la x_{\ell} y_{\ell+\tau}^* x_m^* y_{m+\ups}\ra 
=\la x_{\ell}y_{\ell+\tau}^*\ra \la x_{m}^* y_{m+\ups}\ra
+\la x_{\ell}x_{m}^*\ra \la y_{\ell+\tau}^* y_{m+\ups}\ra .
\label{continuous_fourth_moment}
\end{equation}
A third product of second moments,
$\la x_{\ell}y_{m+\ups}\ra \la y_{\ell+\tau}^* x_{m}^*\ra$,
would ordinarily appear on the right-hand side of 
Eq.\ \ref{continuous_fourth_moment},
but vanishes here
because of the assumption that $x$ and $y$ have
no intrinsic phase (Eq.\ \ref{no_intrinsic_phase}).
Eq.\ \ref{rtau_atau_def} gives the second moments,
so that
Eq.\ \ref{rtru_sum_expand} becomes:
\begin{equation}
\la r_{\tau} r_{\ups}^*\ra = \rho_{\tau}\rho_{\ups} + {{1}\over{N_o}}\sum_{n=1}^{N_o}\alp_{n}\alp_{-n+(\tau-\ups)} .
\label{continuous_rtruc}
\end{equation}
Here, I have
used the ``wrap'' assumption for the correlation function (Eq.\ \ref{wrap_assumption}).
The variance is thus:
\begin{equation}
\la r_{\tau} r_{\ups}^*\ra -\la r_{\tau}\ra\la r_{\ups}^*\ra =
{{1}\over{N_o}}\sum_{n=1}^{N_o}\alp_{n}\alp_{-n+(\tau-\ups)} .
\label{continuous_rtruc_stddev}
\end{equation}

Three variances,
or two principal axes and an angle,
are required to fully describe the elliptical distribution of noise in the complex plane.
Because $r_{\tau} r_{\tau}^*$ is always real, we require two more.
A convenient independent statistic is:
\begin{equation}
\la r_{\tau} r_{\ups}\ra -\la r_{\tau}\ra\la r_{\ups}\ra =
{{1}\over{N_o}}\sum_{n=1}^{N_o}\rho_{n}\rho_{-n+(\tau-\ups)} .
\label{continuous_rtru_stddev}
\end{equation}
This expression is, in general, complex and thus provides the needed additional two statistics.
As an example, 
one can easily recover the expressions given in Paper 1 for the noise of a
``white'' spectrum, for continuous-valued data, from Eqs.\ \ref{continuous_rtruc_stddev} and \ref{continuous_rtru_stddev}.

\subsubsection{Noise for Spectrum}\label{continuous_spectrum_noise}

The variances of the spectral channels give the noise.
One can obtain the variance by Fourier transforming Eq.\ \ref{continuous_rtruc}:
\begin{eqnarray}
\la \twr_k \twr_k^*\ra &=& 
\sum_{\tau=-N}^{N-1} \sum_{\ups=-N}^{N-1} e^{i{{2\pi}\over{2N}}k (\tau-\ups)}
\la r_{\tau}r_{\ups}^* \ra \\
&=& \twp_k \twp_k^* + {{2N}\over{N_o}}\twa_k \twa_k . \nonumber
\end{eqnarray}
This uses the fact that the Fourier transform of the autocorrelation function
is the power spectrum
(Eqs.\ \ref{convolution_fact},\ref{convolution_fact_rho}).
I assume here that all nonzero elements of the correlation functions
$\alp_{\tau}$, $\rho_{\tau}$
lie within the range that is transformed to a spectrum, $-N<\tau<N-1$.
In other words, the spectral resolution is sufficient to 
completely resolve all features of the spectrum.
Also, I again use the wrap assumption, Eq.\ \ref{wrap_assumption}.
Thus,
\begin{equation}
\la \twr_k \twr_k^*\ra - \la \twr_k \ra\la \twr_k^*\ra 
= {{2N}\over{N_o}} \twa_k \twa_k^* .  
\label{continuous_twrktwrkc}
\end{equation}
Analogously from Eq.\ \ref{continuous_rtru_stddev} one finds:
\begin{equation}
\la \twr_k \twr_k\ra -\la \twr_k \ra\la \twr_k\ra = {{2N}\over{N_o}} \twp_k \twp_k . 
\label{continuous_twrktwrk}
\end{equation}

Together, Eq. \ref{continuous_twrktwrkc} and \ref{continuous_twrktwrk}
describe the noise of the cross-power spectrum.
Note that the noise, measured as the standard deviation,
increases proportionately with the square root of the number of spectral channels $\sqrt{2N}$,
and decreases as the inverse square root of number of measurements $\sqrt{2 N_o}$.
Each of the $N_o$ complex terms in the correlation function
involves measurement of two quantities, 
so that for counting arguments the number of independent data is actually $2 N_o$.

If we suppose that a particular element $\twp_k$ of the cross-power spectrum is real
(or, equivalently, if we rotate the phase of $x$ until $\twp_k$ is real!),
then Eqs.\ \ref{continuous_twrktwrkc} and\ \ref{continuous_twrktwrk}
show that:
\begin{eqnarray}
\la \re[\twr_k]\ra &=& \twp_k \label{continuous_spectrum_facts} \\
\la \im[\twr_k]\ra &=& 0 \nonumber \\
\la \re[\twr_k]\re[\twr_k]\ra - \la \re[\twr_k]\ra\la\re[\twr_k]\ra &=& {{2N}\over{2N_o}}(|\twa_k|^2 + \twp_k^2) \nonumber \\
\la \im[\twr_k]\im[\twr_k]\ra \phantom{- \la \re[\twr_k]\ra\la\re[\twr_k]\ra }
&=& {{2N}\over{2N_o}}(|\twa_k|^2 - \twp_k^2) \nonumber \\
\la \re[\twr_k]\im[\twr_k]\ra \phantom{- \la \re[\twr_k]\ra\la\re[\twr_k]\ra }
&=& 0 . \nonumber
\end{eqnarray}
These equations describe the error ellipses in the complex plane
for spectral measurements.
They are consistent with the results of Paper 1 for a white spectrum ($\alp_k=1$, $\rho_k={\rm const}$),
and are closely related to ``self-noise'' (see Paper 1).

The noise in the measured autocorrelation spectrum $\twe_k$ is identical to that in
the cross-power spectrum $\twr_k$ (Eq.\ \ref{continuous_twrktwrk} or 
\ref{continuous_spectrum_facts}), with substitution of $\twa_k$ for $\twp_k$.

\subsubsection{Noise is Uncorrelated Between Spectral Channels}\label{noise_uncorr_continuous}

The correlation of noise between spectral channels 
can be found from a generalization of Eq.\ \ref{continuous_twrktwrkc}:
\begin{eqnarray}
\la \twr_k \twr_{\ell}^* \ra &=& \sum_{\tau,\ups=-N}^{N-1} e^{i(k\tau -\ell\ups)} \la r_{\tau} r_{\ups}^* \ra \\
&=&
\la \twr_k\ra\la \twr_{\ell}^*\ra +
{{2 N}\over{2 N_o}} \sum_{\ups,\mu=-N}^{N-1} \sum_{m,n=1}^{N_o} e^{i(k\mu+(k-\ell)\ups)}\la a_{n-m}\ra\la a_{-(n-m)+\mu}\ra \nonumber \\
&=& 0 , \quad {\rm unless}\ \ell=m .
\end{eqnarray}
Here, I have introduced 
$\mu=\tau-\ups$.
The summation over 
$\ups$ 
yields zero unless 
$\ell=m$ 
(in which case one recovers 
Eq.\ \ref{continuous_twrktwrkc}).
Thus,
noise is uncorrelated between different channels, for the spectrum of a continuous signal.

\section{CORRELATION FUNCTIONS OF QUANTIZED SIGNALS}\label{quantized}

\subsection{Quantized Gaussian Noise}

\begin{figure}[t]
\epsscale{.50}
\plotone{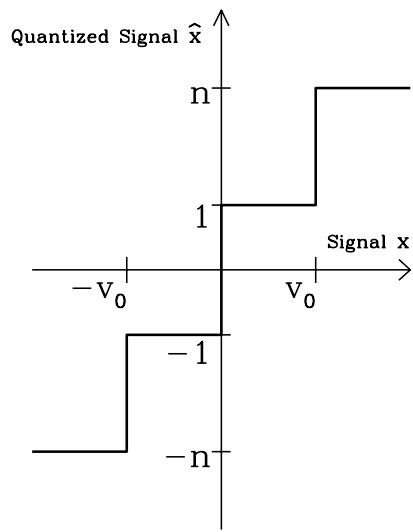} 
\figcaption[]{
Characteristic curve for 4-level quantization.
\label{4_level}}
\end{figure}

Suppose now that the time-series $x_{\ell}$
and $y_{\ell}$ are quantized,
to produce the time series $\hat x_{\ell}$
and $\hat y_{\ell}$.
Quantization involves converting value of the continuous
variables $x_{\ell}$
and $y_{\ell}$ to one of a discrete set of values via a 
characteristic curve.  Fig.\ \ref{4_level} shows an example,
for 4-level quantization.
Such curves can be parametrized by the locations of the steps, $\{v_{xi}\}$ and $\{v_{yi}\}$,
and the weights of each step, $\{n_{i}\}$.
I assume that the 
same curve is used for the real and imaginary parts of 
both $x_{\ell}$ and $y_{\ell}$,
although the curve for $x_{\ell}$ may differ from that for $y_{\ell}$.
I also assume that the characteristic curve is antisymmetric
for both real and imaginary parts:
$\hat X (X) = -\hat X(-X)$,
where $X$ is the real or imaginary part of $x$;
and analogously for $y$.
Paper 1 discusses additional details of quantization, with references.
Quantization will preserve some properties of 
the continuous signals and their correlation functions and spectra,
and change others, as this section investigates.

\subsection{Correlation Function for Quantized Data}

From the quantized time series $x_{\ell}$ and $y_{\ell}$,
one can form the
cross-correlation function $\hat r_{\tau}$, 
\begin{equation}
\hat r_{\tau}={{1}\over{2 N_o}}\sum_{\ell=1}^{N_o}\hat x_{\ell}\, \hat y_{\ell+\tau}^* , \label{rtau_def} \\
\end{equation}
and the autocorrelation function of $\hat x$:
\begin{equation}
\hat a_{\tau}={{1}\over{2 N_o}}\sum_{\ell=1}^{N_o}\hat x_{\ell}\, \hat x_{\ell+\tau}^* .
\label{atau_def}
\end{equation}
Again I use the ``wrap'' assumption, Eq.\ \ref{wrap_assumption}.
Note that $\hat a$ may differ for the series $x$ and $y$ because of differences in characteristic curves,
as well as for reasons noted above.
One seeks to relate $\hat r_{\tau}$ and $\hat a_{\tau}$ as closely as possible to 
the ensemble averages for continuous data, 
$\rho_{\tau}$ and $\alp_{\tau}$, 
via a simple deterministic relationship
and with as little noise as possible.

Among the classic treatments of correlation of quantized signals
are the works of \citet{coo70} and \citet{jen98}.
In the notation of Paper 1 and the following sections,
Cooper found that $\hat r(\rho)$ is proportional to $\rho$, for small $\rho$,
and determined the constant of proportionality.
\citet{jen98} pointed out that this proportionality is 
quite accurate until $\rho$ approaches 1 closely, where the departure becomes significant.
Most cross-correlations of astrophysical data yield small $\rho$,
justifying the linear approximation.
However, for autocorrelation, the
``zero lag'' must yield unit correlation: $\alp_0=1$ 
(see \S\ref{continuous_avg} above),
for which the linear approximation is poor.
Jenet \& Anderson concluded 
 that the autocorrelation function for quantized data is nearly proportional
to the desired result $\alp_{\tau}$, with an additional spike
at zero lag.

\subsection{Simulations of Cross-Correlation}\label{simulations}

\begin{figure}[t]
\epsscale{.60}
\plotone{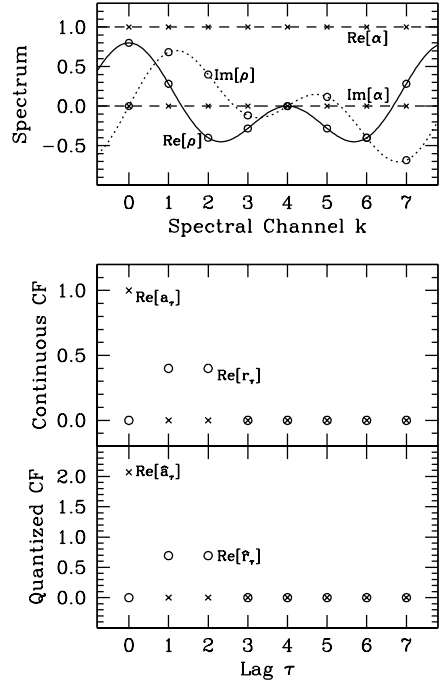} 
\figcaption[]{
Model spectra and correlation functions for simulations of correlation.
Upper panel: Cross-power spectrum $\twr_k$ (circles) 
and autocorrelation spectrum $\twa_k$ (crosses).
Curves show interpolated spectrum.
Middle panel: Cross-correlation function $r_{\tau}$ and autocorrelation function $a_{\tau}$
for continuous data.
Lower panel: Cross-correlation function $\hat r_{\tau}$ and autocorrelation function $\hat a_{\tau}$
for quantized data. Note the gain for 
cross-correlation of $\hat r_{\tau} =B_x B_y 0.4 = 0.693$ for $\tau=1,2$,
and offset to $a_0 = 2.07$ for the zero-lag autocorrelation.
Data were quantized with 
$v_0=1.5$ and 
$n=3$. Correlation includes $2N=8$ lags.
When
averaged over $N_o =2\times 10^6$ simulated correlation functions,
the simulated spectra and correlation functions are indistinguishable from theoretical values.
\label{dxcf_avgplot}}
\end{figure}

For comparison with analytical results, I simulated correlation 
of Gaussian noise.
Figure\ \ref{dxcf_avgplot}
shows the average spectra and correlation functions for one simulation,
with $2N=8$ lags, used as an example in the rest of the paper.
The autocorrelation function is 
``white'' with $\alpha_{\tau}=1$ for $\tau = 0$, and $\alpha_{\tau}=0$ for $\tau\neq 0$.
The cross-correlation function has 
only 2 nonzero lags, $\tau=1,2$: $\rho_1=\rho_2=0.4$.
Note that this is somewhat different from typical radioastronomical
data, which typically contain a white background noise
spectrum (which appears as a spike in the autocorrelation function
at $\tau=0$),
with an admixture of spectrally-varying noise,
perhaps with varying correlation.

I formed the original noiselike data for Figure\ \ref{dxcf_avgplot} by 
drawing elements from Gaussian distributions for each spectral channel.
This method reflects the underlying assumption that the
spectrum consists of a number of independent spectral components
with different frequencies.
For each spectral channel, the Gaussian distribution had unit variance
(as indicated by the flat autocorrelation spectrum $\alpha_k = 1$ in the 
upper panel of Figure\ \ref{dxcf_avgplot}).
However, correlations between the conjugates of $x_k$ and $y_k$ varied with
spectral channel $k$, to yield the spectral variation of $\rho_k$ seen in the figure.
Paper 1 (\S\ 4) describes formation of such a distribution.  
For this work, the phase of one series was rotated, in each channel, to produce the 
phase desired for $\rho_k$.
I then Fourier transformed these frequency-domain data to the time domain,
to produce the series 
$x_{\ell}$ and $y_{\ell}$.
This yielded Gaussian noise with the desired correlations.
I then quantized these series using a characteristic curve as in Figure\ \ref{4_level}
with $v_0=1.5$, $n=3$ to form the series $\hat x_{\ell}$, $\hat y_{\ell}$.
After quantization,
I correlated the 
time series to produce the correlation function
$\hat r_{\tau}$. 
I discuss Fourier transform of $\hat r_{\tau}$ to form the quantized spectrum in 
\S\ref{section_quantspectrum} below.

The predictions of  \citet{coo70} and \citet{jen98} for the average correlation function,
re-derived in the following section, agree with the simulation to much better than the
size of the points in the figure.
In the following sections, I calculate the expected noise in the 
correlation function, 
and compare 
results with simulations of this spectrum.

\subsection{Mean Cross-Correlation Function for Quantized Data}

To introduce the analytical technique used 
to find the noise below,
I re-derive the results of \citet{coo70} and \citet{jen98}.
Eq.\ \ref{rtau_def} gives the ensemble-average autocorrelation function:
\begin{equation}
\la \hat r_{\tau} \ra = {{1}\over{2 N_o}}\sum_{\ell} \la \hxh_{\ell} \hyh_{\ell+\tau}\ra .
\label{digital_rt_expand}
\end{equation}
The quantity $\la \hat x_{\ell} \hat y_{\ell+\tau}^*\ra$
is of the form $\la \hat w \hat x^*\ra$, where $\hat w$ and $\hat x$ are quantized random variables. 
This average can be expanded into products of
pairs of real and imaginary parts of $\hat w$ and $\hat x$:
\begin{equation}
\la \hat w \hat x^*\ra = \big(\la\re[\hat w]\re[\hat x]\ra + \la\im[\hat w]\im[\hat x]\ra\big) 
+ i\big(\la\im[\hat w]\re[\hat x]\ra - \la\re[\hat w]\im[\hat x]\ra\big) .
\label{reimparts_wxc}
\end{equation}
The various averages of the 
{\it real} quantized Gaussian variables on the right-hand side of this
equation are given in Table\ \ref{table_real_avgs};
in this case, by the first line:
$\la\hat W \hat X\ra = B_W B_X \rho_{WX}$.
Here, $W$ and $X$ are real (or imaginary) variables drawn from the bivariate
Gaussian distribution with covariance $\rho_{WX}$,
and $\hat W$ and $\hat X$ are their quantized counterparts.
The statistical average $\la ...\ra$ is an integral over the probability
distribution for $W$ and $X$, times the characteristic curves for $\hat W(W)$ and $\hat X(X)$.
In  \S\ 3.2.1 of Paper 1, this expression was expanded in powers of $\rho_{XY}$ to yield
the term in the second column of Table\ \ref{table_real_avgs},
times one-dimensional Gaussian distributions of $W$
and $X$ and their 
characteristic curves.
Integration over $X$ and $Y$ 
yields the term in the third column in Table\ \ref{table_real_avgs}.

As Eq.\ \ref{reimparts_wxc} shows,
several expressions of the form $\la \hat W(W)\hat X(X)\ra$ must be combined
to find the complex average $\la \hat w \hat x^*\ra$.
The covariances of the various real and imaginary parts
can be combined to form a complex covariance, $\rho_{WX}$:
\begin{eqnarray}
\la\re[w]\re[x]\ra &=& \phantom{-}\la\im[w]\im[x]\ra = \re[\rho_{WX}] \\
\la\im[w]\re[x]\ra &=& -\la\re[w]\im[x]\ra = \im[\rho_{WX}] .  \nonumber
\end{eqnarray}
One thus obtains the expression for $\la \hat w \hat x^*\ra$ given in the first
line of Table\ \ref{table_complex_avgs}, in the third column:
\begin{equation}
\la \hat w \hat x^*\ra = 2 [B_X B_Y] \rho_{WX} .
\label{wxc}
\end{equation}
Note that this result is accurate through second order;
as discussed in Paper 1, the next correction is third-order.
Substitution into Eq.\ \ref{digital_rt_expand} recovers the result of \citet{coo70}, here with complex correlations:
\begin{equation}
\la \hat r_{\tau} \ra = B_X B_Y \rho_{\tau} . \label{avg_rtau_rhotau_prop}
\end{equation}

\subsection{Mean Autocorrelation Function for Quantized Data}

As \citet{jen98} point out, the mean autocorrelation function must be
treated differently from cross-correlation.
Eq.\ \ref{atau_def} gives the ensemble-average autocorrelation function:
\begin{equation}
\la \hat a_{\tau} \ra = {{1}\over{2 N_o}}\sum_{\ell} \la \hxh_{\ell} \hxh_{\ell+\tau}\ra .
\end{equation}
This involves products of different elements for $\tau\neq 0$,
and square moduli of elements for $\tau=0$.
Thus, it involves terms of both the form $\la \hat w \hat x^*\ra$,
and of the form $\la \hat w \hat w^*\ra$.
The first is the same as for cross-correlation;
the second requires a different, though analogous, calculation.
The results in the first 2 lines in Table\ \ref{table_complex_avgs},
yield the expression of \citet{jen98} for the 
statistically-averaged cross-power spectrum:
\begin{eqnarray}
\la \hat a_{\tau}\ra &=&\cases { A_{X2},& if $\tau=0$; \cr
B_X^2 \alp_{\tau}, & if $\tau\neq 0$. \cr}
\label{avg_hat_a}
\end{eqnarray}
Again, the constants $A_{X2}$ and $B_X$ depend on the characteristic curve;
Paper 1 presents expressions for them.
The result holds through second order in $\alp_{\tau}$.
Figure \ref{dxcf_avgplot} illustrates the resulting spike at zero lag, for autocorrelation.

\subsection{Noise of Cross-Correlation Functions for Quantized Signals}

The variance of the correlation function measures the noise.
The noise
thus involves the fourth moment of the quantized signals $\hat x_{\ell}$ and $\hat y_{\ell}$.
Because the correlation function is complex,
it is drawn from an elliptical Gaussian distribution in the complex plane, 
and one must determine both $\la \hat r\hat r^*\ra$
and $\la \hat r\hat r\ra$ to characterize its noise.
Both of these expressions are sums of terms
of the general form $\la \hat w \hat x^* \hat y^* \hat z\ra$, or $\la \hat w \hat x^* \hat y \hat z^*\ra$.
Up to 2 of the 4 quantities $\hat w \hat x \hat y \hat z$ can be
identical 
for the cross-power spectrum,
and all of them can be identical for the autocorrelation spectrum.
The identical quantities result in special cases, for quantized data,
as Jenet \& Anderson found. 

Precisely along the lines of the discussion of the second moments
in the preceding section,
expansion of the fourth moments into real and imaginary parts 
yields statistical averages of the form $\la \hat W \hat X \hat Y \hat Z\ra$,
where $W$ $X$ $Y$ and $Z$ are real quantities drawn from 
a multivariate Gaussian distribution.
The first column of Table\ \ref{table_real_avgs} lists the terms
important for the correlation functions.
I expand the multivariate Gaussian distribution
for $W$ $X$ $Y$ and $Z$ through second order in covariances
$\rho_{WX}$, $\rho_{WY}$, and so on;
this yields the terms in the second column of Table\ \ref{table_real_avgs},
times 1D Gaussian distributions for each variable.
Multiplication by the quantizing functions
$\hat W(W)$ $\hat X(X)$ $\hat Y(Y)$ and $\hat Z(Z)$ and integration over the distributions
yields the averages in the third column of Table\ \ref{table_real_avgs}.
These averages of quantized real (or imaginary) quantities 
combine to yield the averages of quantized complex quantities
given in Table\ \ref{table_complex_avgs}.
I then combine these averages, using the schemes summarized in Table\ \ref{table_terms_in_Xsums}
to find expressions for the variance of the cross-correlation function $\hat r$.

\subsubsection{$\la\hat r\hat r^*\ra - \la\hat r\ra\la\hat r^*\ra$}\label{rtrtc}

The noise in the modulus of the correlation function,
$\la\hat r\hat r^*\ra - \la\hat r\ra\la\hat r^*\ra$,
gives the average diameter of the error ellipse for $\hat r$.
To find this,
one must calculate
\begin{equation}
\la\hat r_\tau\hat r_\ups^*\ra = {{1}\over{(2 N_o)^2}} 
\sum_{\ell, m =1}^{N_o}
\la x_{\ell}y^*_{\ell+\tau}x^*_{m}y_{m+\ups} \ra  .
\label{quant_rtruc}
\end{equation}
Again,
I assume that covariances between terms are small,
so that expansion through second order is sufficient.

The calculation 
is straightforward when
all 4 of the averaged elements are different: 
in other words, when
$\ell\neq m$ and $\ell+\tau\neq m+\ups$.
In this case,
the average is proportional to
that expected for continuous correlation, Eq.\ \ref{continuous_fourth_moment}:
\begin{equation}
\la \hat x_\ell \hat y_{\ell+\tau}^* \hat x_m^* \hat y_{m+\ups} \ra = 
\big[4 B_X^2 B_Y^2\big] \rho_\tau \rho_\ups^* + \big[4 B_X^2 B_Y^2\big] \left( \alp_{m-\ell}\alp_{-(m-\ell)+(\tau-\ups)}\right) .
\label{WXcYZc}
\end{equation}
This is the average given by the term
$\la \hat w \hxc \hyc \hat z \ra$ in Table\ \ref{table_complex_avgs},
where it appears as ``class'' $1111+$.
The 1's indicate that one term of each variable appears once;
the ``$+$'' indicate the symmetry of average under 
multiplication of $x$ by $e^{i\pi/2}$, or equivalently
rotation by $\pi/2$ in the complex plane.
This term also appears in Table\ \ref{table_terms_in_Xsums},
with ID ``Xcn.0''.  
In this identifier,
the ``X'' indicates cross-correlation,
the ``c'' indicates the product of $\hat r$ with its conjugate: $\hat r\hat r^*$,
the ``n'' indicates that $\tau\neq\ups$,
and the ``0'' indicates that the indices $\ell$, $m$, $\ell+\tau$, and $m+\ups$ are distinct.
As the table indicates under ``Multiplicity,'' 
this form of term appears $N_o^2-2 N_o$ times in the sum.

If 
$\tau\neq\ups$, but $\ell=m$, then one encounters the average
\begin{equation}
\la \hat x_\ell \hat y_{\ell+\tau}^* \hat x_\ell^* \hat y_{\ell+\ups} \ra = 
\big[2 (C_{X2}-A_{X2})B_Y^2\big]\rho_{\tau}\rho_{\ups}^* 
+\big[4 A_{X2}B_Y^2\big]\alp_{(\tau-\ups)}.
\label{WXcWcY}
\end{equation}
This term has
the form $\la \hat w \hxc \hwc \hat y \ra$, and ``Class'' $211+$ in
Table\ \ref{table_complex_avgs}.
It appears as ``Xcn.1'' in Table\ \ref{table_terms_in_Xsums},
and appears $N_o$ times in the sum.

If $\tau\neq\ups$, but $\ell+\tau=m+\ups$, one then encounters
\begin{equation}
\la \hat x_\ell \hat y_{\ell+\tau}^* \hat x_{\ell+\tau-\ups}^* \hat y_{\ell+\tau} \ra = 
\big[2 B_X^2(C_{Y2}-A_{Y2})\big]\rho_{\tau}\rho_{\ups}^* +\big[4 B_X^2A_{Y2}\big]\alp_{(\tau-\ups)}\; .
\label{WXcYcX}
\end{equation}
This term also has
the form $\la \hat w \hxc \hwc \hat y \ra$, and Class $211+$ in
Table\ \ref{table_complex_avgs}.
(Note however that the roles of $\hat x$ and $\hat y$ 
are interchanged from those in Table\ \ref{table_complex_avgs}).
It appears as ``Xcn.2'' in Table\ \ref{table_terms_in_Xsums},
and appears $N_o$ times in the sum.

From Eqs.\ \ref{WXcYZc} through\ \ref{WXcYcX}, I evaluate the sum, Eq.\ \ref{quant_rtruc} (for $\tau\neq\ups$):
\begin{eqnarray}
\la \hat r_{\tau} \hat r_{\ups}^*\ra &=& 
{{1}\over{(2 N_o)^2}}\biggl\{
N_o^2 \big[4 B_x^2 B_y^2\big]\rho_\tau\rho_\ups^* 
+N_o \sum_{n=1}^{N_o}\big[4 B_x^2 B_y^2\big]\alp_{-n} \alp_{n+(\tau-\ups)} \label{prolix_digital_rtruc} \\
&&\phantom{{{1}\over{(2 N_o)^2}}}
-2\times N_o \Bigl\{\big[ 4B_x^2 B_y^2\big]\rho_\tau \rho_\ups^* + \big[ 4B_x^2 B_y^2\big]\alp_0 \alp_{(\tau-\ups)} \Bigr\} \nonumber \\
&&\phantom{{{1}\over{(2 N_o)^2}}}
+N_o \left(\big[2(C_{X2}-A_{X2})B_Y^2\big]\rho_{\tau}\rho_{\ups}^* 
+\big[4 A_{X2}B_Y^2\big]\alp_{(\tau-\ups)}\right) \nonumber \\
&&\phantom{{{1}\over{(2 N_o)^2}}}
+N_o \left(\big[2B_X^2(C_{Y2}-A_{Y2})\big]\rho_{\tau}\rho_{\ups}^* +\big[ 4B_X^2A_{Y2}\big]\alp_{(\tau-\ups)}\right) \biggr\} . \nonumber
\end{eqnarray}
Note that the first 2 terms on the right side of this equation 
give the contribution for all unlike $w x y z$, Eq.\ \ref{WXcYZc},
with multiplicity $2 N_o$ greater than correct.
The second 2 terms subtract off the extras for the special cases $\ell=m$ and $\ell+\tau=m+\ups$,
with multiplicity of $N_o$ each;
and the last 4 terms add back in the correct contributions for these 2 special cases
(Eqs.\ \ref{WXcWcY} and\ \ref{WXcYcX}),
with multiplicity $N_o$ each.
Eq.\ \ref{prolix_digital_rtruc} simplifies to:
\begin{eqnarray}
\la\hat r_{\tau}\hat r_{\ups}^*\ra 
&-& \la\hat r_{\tau}\ra\la\hat r_{\ups}^*\ra =
{{1}\over{2 N_o}}\sum_{n=1}^{N_o}\big[2B_X^2 B_Y^2\big]\alp_{n+(\tau-\ups)}\alp_{-n} \label{digital_rtruc}  \\
&&+{{1}\over{2 N_o}}\big[(C_{X2}-A_{X2})B_Y^2+B_X^2(C_{Y2}-A_{Y2})-4B_X^2 B_Y^2\big]\rho_{\tau}\rho_{\ups}^* \nonumber \\
&&+{{1}\over{2 N_o}}\big[2A_{X2}B_Y^2+2B_X^2A_{Y2}-4B_X^2 B_Y^2\big]\alp_{(\tau-\ups)} . \nonumber
\end{eqnarray}

Similarly, when $\tau=\ups$, the contributing terms are given under  Xce in 
Table\ \ref{table_terms_in_Xsums}.
The case $\ell=m$ again presents a special situation;
for $\tau=\ups$ this case is identical to $\ell+\tau = m+\ups$.
With this special case $\ell=m$ again included incorrectly,
subtracted back off, and then added in correctly, one finds:
\begin{eqnarray}
\la \hat r_{\tau} \hat r_{\tau}^*\ra
 &-& \la\hat r_{\tau}\ra\la\hat r_{\tau}^*\ra =
{{1}\over{2 N_o}}\sum_{n=1}^{N_o}\big[2 B_X^2 B_Y^2\big]\alp_n\alp_{-n} \label{digital_rtrtc} \\
&&+{{1}\over{2 N_o}}\big[\haf(C_{X2}-A_{X2})(C_{Y2}-A_{Y2})-2B_X^2 B_Y^2\big]\rho_{\tau}\rho_{\tau}^* \nonumber \\
&&+{{1}\over{2 N_o}}\big[2A_{X2}A_{Y2}-2B_X^2 B_Y^2\big] \nonumber
\end{eqnarray}

\subsubsection{$\la\hat r\hat r\ra - \la\hat r\ra\la\hat r\ra$}\label{rtrt}

The variance of the correlation function,
given by $\la\hat r \hat r\ra - \la\hat r\ra\la\hat r\ra$,
measures the departure of the error ellipse for $\hat r$ from circularity.
As in the previous section, 
the averages for which 2 or more of the elements of the sum are identical
must be calculated separately.
For $\tau\neq\ups$, the terms appear under Xrn in Table\ \ref{table_terms_in_Xsums}.
This yields:
\begin{eqnarray}
\la\hat r_{\tau}\hat r_{\ups}\ra 
&-& \la\hat r_{\tau}\ra\la\hat r_{\ups}\ra =
{{1}\over{2 N_o}}\sum_{n=1}^{N_o} \big[2B_X^2 B_Y^2\big]\rho_{n+(\tau+\nu)}\rho_{-n} \label{digital_rtru} \\
&&+{{1}\over{2 N_o}}\big[(C_{X2}-A_{X2})B_Y^2+B_X^2(C_{Y2}-A_{Y2})-4B_X^2 B_Y^2\big]\rho_{\tau}\rho_{\ups} \; . \nonumber
\end{eqnarray}
Similarly for $\tau=\ups$,
for which the terms appear under Xre in Table\ \ref{table_terms_in_Xsums}:
\begin{eqnarray}
\la\hat r_{\tau}\hat r_{\tau}\ra 
&-& \la\hat r_{\tau}\ra\la\hat r_{\tau}\ra  = 
{{1}\over{2 N_o}}\sum_{n=1}^{N_o} \big[2B_X^2 B_Y^2\big]\rho_{n+(2\tau)}\rho_{-n} \label{digital_rtrt} \\
&&+{{1}\over{2 N_o}}\big[(\haf(C_{X2}-A_{X2})+B_X^2)(\haf(C_{Y2}-A_{Y2})+B_Y^2)-4B_X^2 B_Y^2\big]\rho_{\tau}\rho_{\tau} \nonumber\\
&&+{{1}\over{2 N_o}}\big[(\haf(C_{X2}-A_{X2})-B_X^2)(\haf(C_{Y2}-A_{Y2})-B_Y^2)\big]\rho_{\tau}^*\rho_{\tau}^* . \nonumber
\end{eqnarray}

\subsubsection{Simulation of Cross-Correlation Function}\label{xcf_simulate}

\begin{figure}[t]
\epsscale{.90}
\plotone{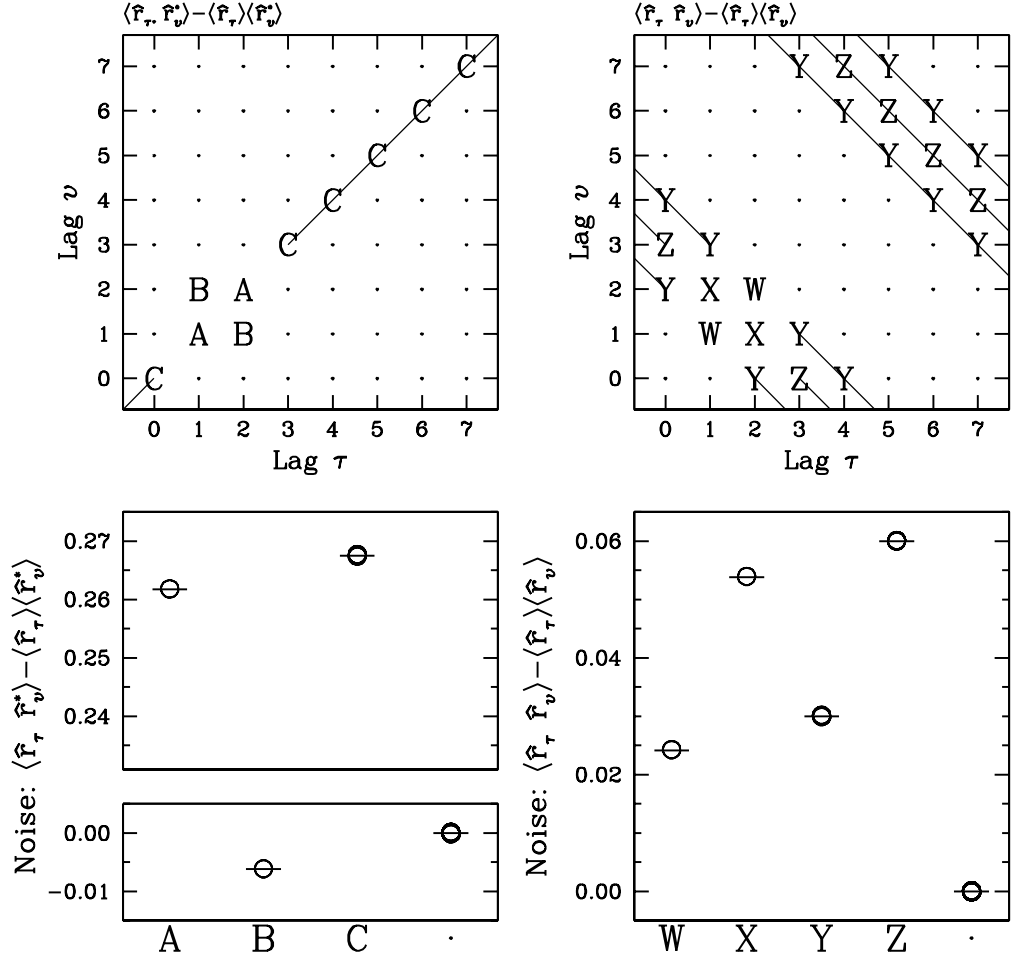} 
\figcaption[]{
Noise for cross-correlation function shown in Figure\ \ref{dxcf_avgplot}.
Upper panels: Schematic depiction of the correlation matrices
$\la r_{\tau} r_{\ups}^*\ra - \la r_{\tau}\ra\la r_{\ups}^*\ra$
(left panel: Equations\ \ref{digital_rtruc} and\ \ref{digital_rtrtc});
and 
$\la r_{\tau} r_{\ups}\ra - \la r_{\tau}\ra\la r_{\ups}\ra$
(right panel: Equations\ \ref{digital_rtru} and\ \ref{digital_rtrt}).
Letters indicate positions with expected nonzero standard deviation,
according to those equations,
with the same standard deviation expected
for identical letters. 
Lower panels: corresponding noise, as found
for a 4-level correlator for the spectrum of
Figure\ \ref{dxcf_avgplot}.
Standard deviations are for $N_o=16$ measurements with $2N=8$ lags,
calculated over $10^6$ simulated correlation functions.
Circles show statistics of the simulations,
and horizontal bars show predictions of 
Eqs.\ \ref{digital_rtruc},\ \ref{digital_rtrtc},
\ \ref{digital_rtru}, and\ \ref{digital_rtrt}.
\label{dxcf_noiseplot}}
\end{figure}

Figure\ \ref{dxcf_noiseplot}
shows statistics,
in the lag domain,
for
the simple correlation function shown in
Figure\ \ref{dxcf_avgplot}.
Plots on the left show
$\la\hat r_{\tau}\hat r_{\ups}^*\ra
-\la\hat r_{\tau}\ra\la\hat r_{\ups}^*\ra$,
and on the right 
$\la\hat r_{\tau}\hat r_{\ups}\ra
-\la\hat r_{\tau}\ra\la\hat r_{\ups}\ra$ .
The upper plot shows
the arrangement of nonzero terms,
and the lower plot gives their values.

The diagonal terms are the 
squared standard deviations
of the amplitude
of $\hat r_{\tau}$,
as given by Eq.\ \ref{digital_rtrtc}.
The off-diagonal terms give the covariances
of the noise between lags,
as given by Eq.\ \ref{digital_rtruc}.

The right panels show the moments
$\la\hat r_{\tau}\hat r_{\tau}\ra
-\la\hat r_{\tau}\ra\la\hat r_{\tau}\ra$.
For a real cross-correlation function
(like that used here),
the diagonal terms are the {\it differences}
of the standard deviations of real and imaginary
parts of $\hat r_{\tau}$,
as given by
Eq.\ \ref{digital_rtrt}.
They thus measure the departure
of the noise from isotropy in phase.
These terms are proportional to squares or products of the cross-correlation function $\rho$.
For this test data, $\rho^2=0.16$, and so these terms are smaller than the largest terms in the left panels.
This indicates that the error ellipses for the correlation function are approximately circular.

\subsection{Autocorrelation Functions}\label{acf_subsection}

Autocorrelation correlation functions and spectra present many special cases.
On the other hand,
for the autocorrelations 
the ``zero lags'' $\tau=0$ and $\ups=0$
yield unit correlation, and thus play a special role;
this is unlike the cross-correlations,
where the quantities being correlated
are distinct at any lag.
Fortunately, one needs only one of $\la\hah_{\tau}\hac_{\ups}\ra$
and $\la\hah_{\tau}\hah_{\ups}\ra$
because 
$\la\hah_{\tau}\hah_{\ups}\ra =\la\hah_{\tau}\hac_{-\ups}\ra$.
Furthermore, $X$ and $Y$ are the same,
so I simplify the notation by dropping the subscripts from the integrals $A$, $B$, $C$.

For the case
$\tau\neq\ups$, 
we have the the general case where neither $\tau$ nor $\ups$ is $0$, as well as the special sub-cases $\tau=0$ and $\ups=0$.
Table\ \ref{table_terms_in_Asums} summarizes these various
cases,
with identifiers Antu, An0u and Ant0.
In these identifiers,
``A'' indicates autocorrelation,
``n'' indicates $\tau\neq\ups$,
and ``0u'' indicates $\tau=0$ whereas ``t0'' indicates $\ups=0$.
Within these cases we have the same special
cases as for the cross-correlation function
$\ell=m$ and $\ell+\tau=m+\ups$,
plus the special cases $\ell+\tau=m$
and $\ell=m+\ups$, which are special cases for autocorrelation
(although not for cross-correlation).
These are listed as Antu.1, Antu.2, etc.
Some of these special cases become degenerate when
$\tau=0$ or $\ups=0$.

I adopt the previous strategy of 
subtracting off, and then adding back in, contributions for the special cases.
For autocorrelations with $\tau\neq\ups$, 
and both $\tau\neq 0$ and $\ups\neq 0$,
this requires 
the ``Antu'' terms in Table\ \ref{table_terms_in_Asums}.
The sum simplifies to:
\begin{eqnarray}
\la\hah_{\tau}\hac_{\ups}\ra &-& \la\hah_{\tau}\ra\la\hac_{\ups}\ra =
{{1}\over{2N_o}} [2B^4]\sum_{n=1}^{N_o} \alp_n \alp_{-n+(\tau-\ups)} \label{A1_sum} \\         \label{auto_1}
&&+{{1}\over{2N_o}}[4(C-A)B^2-8 B^4]\alp_{\tau}\alp_{-\ups}                                                        \label{auto_2}
+{{1}\over{2N_o}}[4 A B^2 -4 B^4]\alp_{\tau-\ups} . \nonumber                                                      \label{auto_3}
\end{eqnarray}
Here I have defined $n=\ell-m$. Note that $\ell-m$ takes on different values in the sub-cases
Antu.3 or Antu.4, as compared with Antu.1 or Antu.2, so that the correction terms are different.
This equation is analogous to, but different from, Eq.\ \ref{digital_rtruc}, with which it should be compared.

In the case $\ups=0$, $\tau\neq\ups$ (Ant0 in Table\ \ref{table_terms_in_Asums}), one obtains:
\begin{eqnarray}
\la \hah_{\tau} \hah_0\ra & - & \la\hah_{\tau}\ra\la\hah_{0}\ra =
{{1}\over{2N_o}} [(C-A)B^2]\sum_{n=1}^{N_o} (\alp_{(n+\tau)} \alp_{-n}) \label{ata0c} \\        \label{auto_4A}
&&+{{1}\over{2 N_o}}[2 B_3 B - 2 C B^2]\alp_{\tau} . \nonumber                                                 \label{auto_5A}
\end{eqnarray}
Note here that $B=\int dX\, X e^{-{{1}\over{2}}X^2} \hat X(X)$, whereas 
$B_3=\int dX\, X e^{-{{1}\over{2}}X^2} (\hat X(X))^3$. (See Paper 1.)

One obtains the analogous expression in the case $\tau=0$, $\tau\neq\ups$ (An0u in Table\ \ref{table_terms_in_Asums}).

In the case $\tau=\ups$, $\tau\neq 0$ (Aet in Table\ \ref{table_terms_in_Asums}), one obtains:
\begin{eqnarray}
\la \hah_{\tau} \hac_{\tau}\ra &-& \la\hah_{\tau}\ra\la\hac_{\tau}\ra = 
{{1}\over{2 N_o}} [2 B^4] \sum_n (\alp_n \alp_{-n}) \\                                                                         \label{auto_6}
&& + {{1}\over {2 N_o}} [2 A^2 - 2 B^4]                                                                                                 \label{auto_7}
+ {{1}\over {2 N_o}}[ (\haf)((C-A)+2 B^2)^2\,-\,8 B^4 ] (\alp_{\tau}\alp_{\tau}^*) .  \nonumber    \label{auto_8}
\end{eqnarray}
This equation is analogous to Eq.\ \ref{digital_rtrtc}.
Finally, in the case $\tau=\ups=0$ (Ae0 in Table\ \ref{table_terms_in_Asums}), one obtains:
\begin{eqnarray}
\la \hah_0 \hac_0\ra &-& \la \hah_0\ra\la \hac_0\ra = 
{{1}\over{2 N_o}}[\haf (C-A)^2]\sum_n (\alp_n \alp_{-n}) \label{a0a0c} \\                                      \label{auto_9}
&& +{{1}\over{2 N_o}}[ A_4 - 2 A^2 - \haf(C-A)^2] . \nonumber                                                      \label{auto_10}
\end{eqnarray}
Note here that $A=\int dX\, X e^{-{{1}\over{2}}X^2} (\hat X(X))^2$,
whereas $A_4=\int dX\, X e^{-{{1}\over{2}}X^2} (\hat X(X))^4$.

\section{SPECTRA OF QUANTIZED SIGNALS}\label{section_quantspectrum}

The measured spectrum  
is the Fourier transform of the measured correlation function.
Thus, for quantized data, the cross-power spectrum $\ckr$ and the autocorrelation spectrum $\cke$ are:
\begin{eqnarray}
\ckr_k &=& \sum_{\tau=-N}^{N-1} e^{i{{2\pi}\over{2 N}}k\tau} \hat r_{\tau} . \\
\cke_k &=& \sum_{\tau=-N}^{N-1} e^{i{{2\pi}\over{2 N}}k\tau} \hat a_{\tau} . \nonumber
\end{eqnarray}
\citet{jen98} show that the proportionality factor found by \citet{coo70} relates 
the average of the quantized cross-power spectrum $\la \ckr \ra$ 
to the true cross-power spectrum $\twp$;
and the same factor, 
with an offset resulting from the spike at zero lag, relates $\la \cke \ra$ to $\twa$.

Noise in the spectrum is related to noise in the autocorrelation function by a double Fourier transform.
I use this fact to find the noise in the spectrum, through second order in $\alp$ and $\rho$, in
this section.  I find that many of the terms for noise in the correlation functions are diluted over the channels of the spectrum.  They can be neglected, in many cases, for spectra containing many
channels.  I find that the dominant terms for noise in individual channels of the spectra are
analogous to results for continuous spectra,
given by Eqs.\ \ref{continuous_twrktwrkc} and \ref{continuous_twrktwrk}.
I also find that the noise is correlated between channels.
This is opposite the conclusion for continuous data (\S\ref{noise_uncorr_continuous}).

\subsection{Mean Spectra for Quantized Signals}\label{spectra_mean}

The Fourier transform of the proportionality 
Eq.\ \ref{avg_rtau_rhotau_prop}
yields the ensemble-averaged spectrum:
\begin{equation}
\la \ckr_k \ra = B_X B_Y \twp_k ,
\label{avg_hat_rk}
\end{equation}
where both sides of the expression are complex.

The ensemble average of the Fourier transform of the quantized autocorrelation function is:
\begin{equation}
\la \twe_k\ra = \sum_{\tau=-N}^{N-1} e^{i{{2\pi}\over{2 N}}k\tau} \hat a_{\tau} .
\end{equation}
This sum contains $2N-1$ terms involving $\hat a_{\tau} = B_X^2 \alp_{\tau}$, 
and one involving $\hat a_{0} = A_{X2}$.
I adopt the approach, as in calculations of noise,
of including an incorrect zero-lag term will all others in the sum,
subtracting that incorrect term, 
and then adding the correct term:
\begin{eqnarray}
\la \twe_k\ra &=& \bigg( \sum_{\tau=-N}^{N-1} e^{i{{2\pi}\over{2 N}}k\tau} B_X^2 \alp_{\tau} \bigg)
\; - \bigg( B_X^2 \alp_0 \bigg) 
\; + \bigg( A_{X2} \bigg) . \label{avg_hat_ak} \\
&=& B_X^2\left(\twa_k + \left({{A_{X2}}\over{B_X^2}}-1\right)\right) .  \nonumber
\end{eqnarray}
This recovers the results of \citet{jen98}, who showed that the mean spectrum 
for quantized data is equal to the statistically-averaged spectrum for continuous data,
plus an offset, times the gain factor $B_{X}^2$.

\subsection{Spectral Noise for Quantized Signals}\label{spectral_noise}

\subsubsection{Variances: $\la\ckr_k\ckr^*_k\ra$}

Calculation of the noise in the spectrum involves the Fourier transform of the variance-covariance matrix.
The Appendix summarizes facts useful for this transform.
The approach is analogous to that taken in \S\ref{continuous_spectrum_noise},
via a double Fourier transform.
I use the facts in the Appendix,
together with Eqs.\ \ref{digital_rtruc} and\ \ref{digital_rtrtc}
to find:
\begin{eqnarray}
\la \ckr_k \ckr_k^*\ra &-&
\la \ckr_k\ra\la\ckr_k^*\ra =
{{(2N)}\over{2 N_o}} 
\big[2(A_{X2}+B_X^2(\twa_k-1))(A_{Y2}+B_Y^2(\twa_k-1))\big] \label{digital_ckrkckrkc} \\
&&+{{1}\over{2 N_o}} 
\big[\haf(C_{X2}-A_{X2})B_Y^2+B_X^2\haf(C_{Y2}-A_{Y2})-2B_X^2 B_Y^2\big] 
\twp_k\twp_k^* \nonumber \\
&&\,-\,{{1}\over{2 N_o}}
\big[2(\haf(C_{X2}-A_{X2})-\haf B_X^2)(\haf(C_{Y2}-A_{Y2})-\haf B_Y^2)-\haf B_X^2 B_Y^2\big]
\sum_{\ell=-N}^{N-1} {{1}\over{(2N)}} \twp_\ell\twp_\ell^* . \nonumber
\end{eqnarray}
Note that the first term on the right-hand side is of order $2N$;
the second is of order $1$;
and the third is of order $1/2N$.

\subsubsection{Variances: $\la\ckr_k\ckr_k\ra$}

Using the expressions in the Appendix
together with Eqs.\ \ref{digital_rtru} and\ \ref{digital_rtrt},
I find:
\begin{eqnarray}
\la \ckr_k \ckr_k\ra &-& \la\ckr_k\ra\la\ckr_k\ra = 
+{{(2N)}\over{(2N_o)}}\big[2B_X^2 B_Y^2\big]\twp_k\twp_k \label{digital_ckrkckrk} \\
&&+{{1}\over{(2N_o)}}\big[(C_{X2}-A_{X2})B_Y^2+B_X^2(C_{Y2}-A_{Y2})-4B_X^2B_Y^2\big]
\twp_k\twp_k \nonumber \\
&&+{{1}\over{(2N_o)}}\big[(\haf(C_{X2}-A_{X2})-B_X^2)(\haf(C_{Y2}-A_{Y2})-B_Y^2)\big] 
{{1}\over{(2N)}} \left( \tilde C_{\rho}(k) + \tilde C_{\rho}^*(-k) \right) . \nonumber 
\end{eqnarray}
Again, the first term on the right-hand side is of order $2N$,
the second of order $1$, and the third of order $1/2N$.

\subsection{Correlation of Noise Across Spectral Channels}

For quantized data, noise in different spectral channels can be correlated.
The correlation of noise between channels involves $\la \ckr_k \ckr_{\ell}^* \ra$, with $k\neq\ell$.
These covariances can be
calculated by the double Fourier transform of Eqs.\ \ref{digital_rtruc} and\ \ref{digital_rtrtc}.

\subsubsection{Covariances: $\la\ckr_k\ckr_{\ell}^*\ra$}\label{covar_ckrckrc}

For calculation of covariances between channels, 
classification of the terms in Eq.\ \ref{digital_rtruc} and\ \ref{digital_rtrtc}
is helpful.
In Eq.\ \ref{digital_rtruc}, the first term on the right-hand
side is proportional to the autocorrelation function $\alpha$ convolved with itself,
the second is proportional to the square of the cross-correlation function $\rho$,
and the third is proportional to $\alpha$.
Of these, only the second will contribute to 
the covariance between channels.
None of the 3 terms on the right-hand side of Eq.\ \ref{digital_rtrtc} contribute either,
for $k\neq\ell$.
Thus, only the second term on the right-hand side of Eq.\ \ref{digital_rtruc}
contributes, and it contributes in a simple way:
\begin{equation}
\sum_{\tau=-N}^{N-1}\sum_{\ups=-N}^{N-1}
e^{i{{2\pi}\over{2N}}(k\tau-\ell\ups)}\rho_{\tau}\rho_{\ups}^* =
\twp_k\twp_{\ell}^* ,
\end{equation}
so that
\begin{equation}
\la\ckr_k\ckr_{\ell}^*\ra -\la\ckr_k\ra\la\ckr_{\ell}^*\ra 
= 
{{1}\over{2 N_o}}\big[(C_{X2}-A_{X2})B_Y^2+B_X^2(C_{Y2}-A_{Y2})-4B_X^2 B_Y^2\big]
\twp_k\twp_{\ell}^* ,\quad  {\rm for}\ k\neq\ell .
\label{covar_pkplc}
\end{equation}
The combination of constants 
$\big[(C_{X2}-A_{X2})B_Y^2+B_X^2(C_{Y2}-A_{Y2})-4B_X^2 B_Y^2\big]$
is always less than 0
for $n=3$ (although it can be positive for other values of
$n$), so the covariance is negative in that case.
In other words, when noise increases the height of one
spectral peak, noise will tend to reduce the heights of
other spectral peaks.
Note that the contribution of $\twp_k\twp_k^*$ to the variance
appears in the covariance as well:
this contribution to the noise is perfectly correlated between spectral channels.

\subsubsection{Covariances: $\la\ckr_k\ckr_{\ell}\ra$}\label{covar_ckrckr}

The covariances $\la\ckr_k\ckr_{\ell}\ra$ can be found from 
Eqs.\ \ref{digital_rtru} and\ \ref{digital_rtrt}.
As in the preceding section, 
classification of the terms in
Eqs.\ \ref{digital_rtru} and\ \ref{digital_rtrt} is helpful.
In both expressions,
the first term is proportional to the convolution of the 
cross-power spectrum with itself;
it does not contribute to the covariance.
The other terms in Eq.\ \ref{digital_rtrt} also contribute nothing.
We thus obtain:
\begin{equation}
\la\ckr_k\ckr_{\ell}\ra  -\la\ckr_k\ra\la\ckr_{\ell}\ra = 
{{1}\over{2 N_o}}\big[(C_{X2}-A_{X2})B_Y^2+B_X^2(C_{Y2}-A_{Y2})-4B_X^2 B_Y^2\big]
\twp_k\twp_{\ell} ,\quad k\neq\ell .
\label{covar_pkpl}
\end{equation}
The covariances have the same coefficient for variances  $\la\ckr_k\ckr^*_{\ell}\ra$  
and  $\la\ckr_k\ckr_{\ell}\ra$ .

\subsubsection{Simulation of Cross-Power Spectrum}\label{xps_simulate}

\begin{figure}[t]
\epsscale{.90}
\plotone{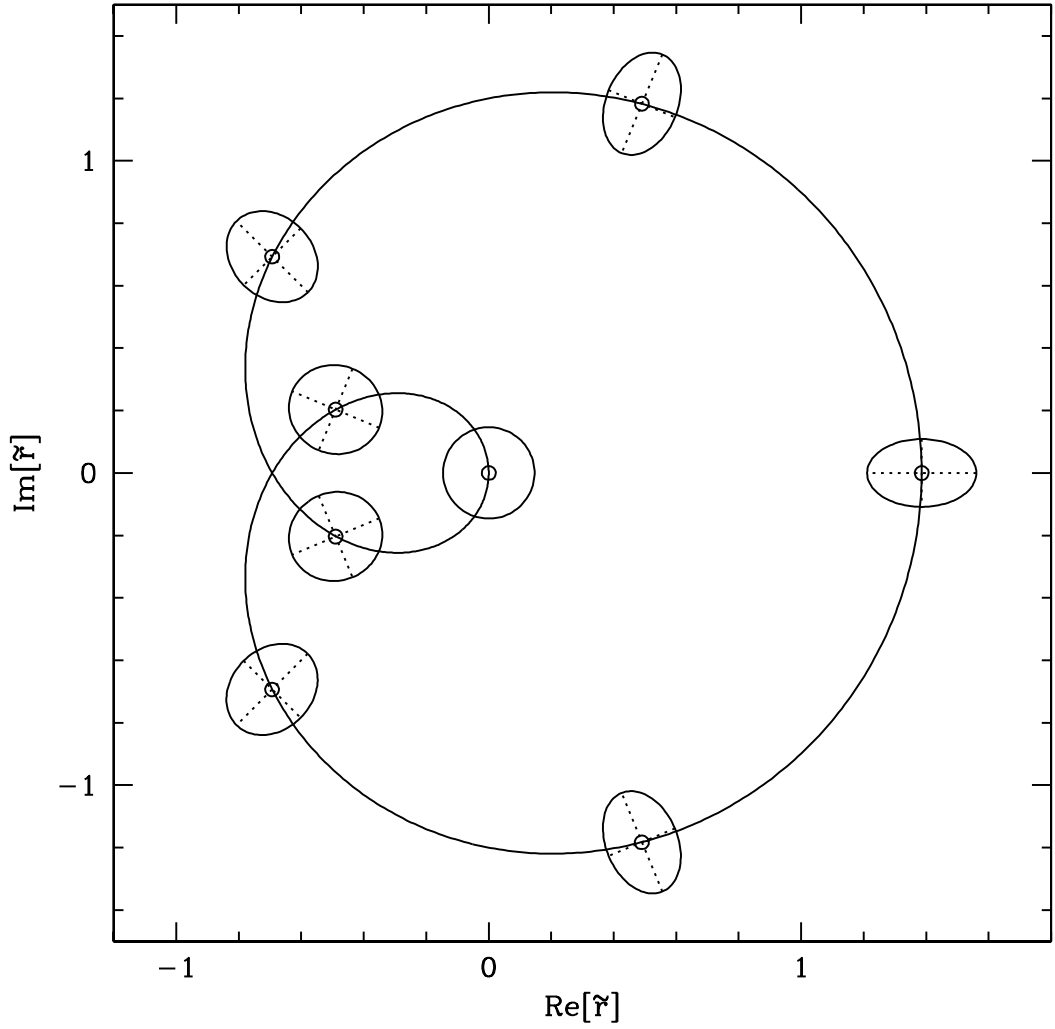} 
%
%
%
\figcaption[]{
Spectrum for model correlation function
of Figures \ref{dxcf_avgplot} and \ref{dxcf_noiseplot},
in phasor form.
Solid line shows expected form,
using Fourier interpolation.
Ellipses show measured averages and standard deviations.
Simulations used a 4-level correlator with
$v_0=1.5$, $n=3$ with $N_o=16$, $2N=8$.
The displayed statistics were calculated from $10^7$ simulated spectra.
The value of $v_0$ was chosen to emphasize the eccentricity of the error 
ellipses;
in other words, of the size of the term $\la \ckr_k \ckr_k^*\ra$.
Note that major axes point toward the origin.
\label{phasor_plot}}
\end{figure}

I Fourier transformed each of the simulated correlation
functions from the simulation and form spectra.
The statistical properties of these spectra are in
good agreement with the results of \S\ref{spectral_noise}.
Figure\ \ref{phasor_plot} shows an example spectrum as
a phasor plot.
This is the spectrum
corresponding to the correlation function
of Figures\ \ref{dxcf_avgplot} and\ \ref{dxcf_noiseplot},
plotted in phasor form.
The prediction is plotted as a solid line,
using Fourier interpolation.
The mean measurements in the discrete channels are plotted as points,
and surrounded by error ellipses that give the spread.
The error ellipses for each point have major axes that point toward
the origin of the complex plane;
this is a consequence of the fact that,
for this choice of parameters,
the first term on the right-hand side of Eq.\ \ref{digital_ckrkckrk}
dominates the other 2,
and it is proportional to $\rho_k^2$.
This term defines the major axis.

\begin{figure}[t]
\epsscale{.80}
\plotone{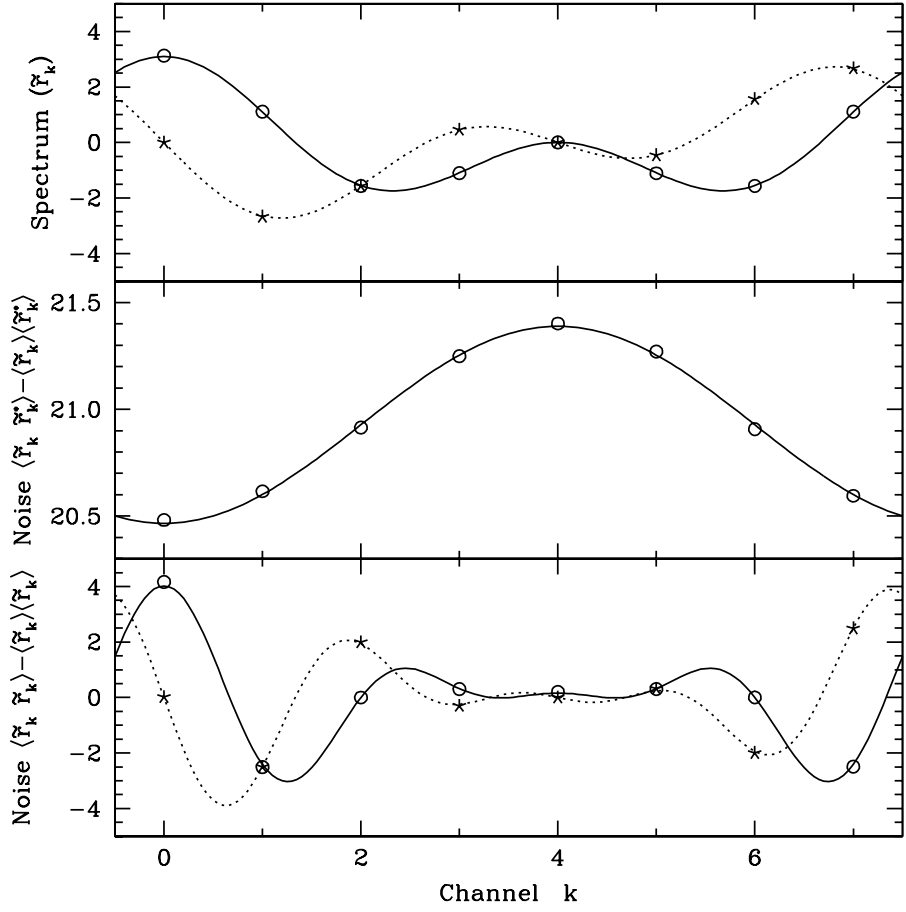} 
\figcaption[]{
Model spectrum and noise, shown for simulations (circles: real parts,
stars: imaginary parts), and for theory (solid line: real part, dotted
line: imaginary part).  Simulations used a 4-level correlator with
$v_0=0.4$, $n=3$ with $N_o=16$, $2N=8$, and $10^7$ simulations.  
The
level $v_0$ was chosen to emphasize the $\rho$-dependent term in 
Eq.\ \ref{digital_ckrkckrkc},
which appears as the variation from a constant value in the middle panel.
\label{noisecompare}}
\end{figure}

Figure\ \ref{noisecompare}
shows a spectrum and the standard deviations plotted in more traditional
form.
Again, I use Fourier interpolation to show
the model as a continuous function of the channel index, $k$.

\subsubsection{Spectrally-Correlated Noise: Simulation}

\begin{figure}[t]
\epsscale{.70}
\plotone{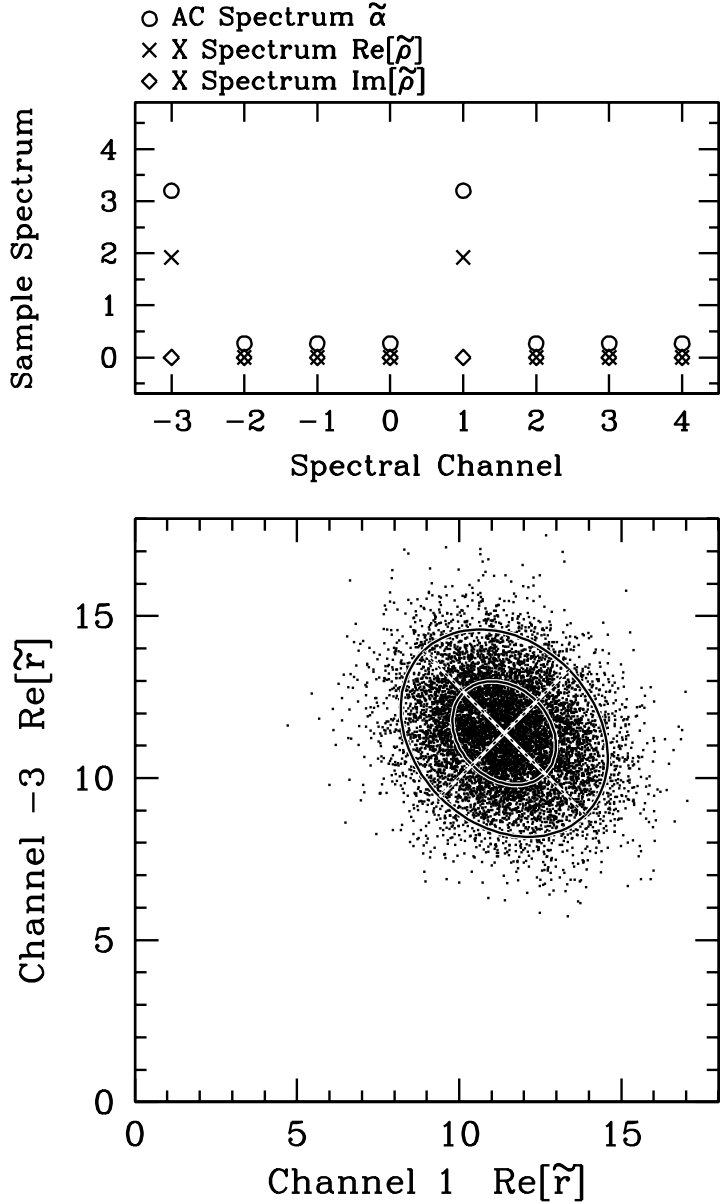} 
\figcaption[]{
Covariance of noise 
in two spectral channels, for simulated correlation of quantized noise.
Upper panel: Model spectrum, showing autocorrelation spectrum $\alpha$ 
and cross-power spectrum $\rho$.
All of the cross-power, and 80\% of the autocorrelation spectrum,
is concentrated into channels $-3$ and $+1$.
Lower panel: Distribution of noise in channels 1 and -3, 
realized from the model spectrum after quantization.
Quantizer parameters were $v_0=0.1$, $n=3$.
Simulations used $N_o=800$ measurements, $2N=8$ spectral channels.
Points show $10^4$ realizations.
Ellipses show 1- and 2-standard deviation contours.
The tilt of the ellipse shows the 
negative covariance of the noise in the two channels.
\label{correl}}
\end{figure}

Figure\ \ref{correl} shows an example of correlated noise in
two spectral channels.
For this simulation I used a different initial spectrum and correlator parameters,
more suited to showing the covariance.
Both channels have strong signals, with zero phase, 
as the spectrum in the upper panel shows.
The lower panel shows results of simulations of correlation of quantized data.
The mean values $\re[\la\ckr_k]$ and $\re[\ckr_{\ell}\ra]$
locate the centroid of the ellipse.
Noise gives the ellipse extension.
The covariance of noise tilts the ellipse:
when $\ckr_k$ is smaller than its mean,
$\ckr_{\ell}$ tends to be larger; and vice versa.
This demonstrates the correlation of noise between two channels.

Comparison of Eqs.\ \ref{covar_pkplc} and\ \ref{covar_pkpl}
shows that the correlated noise is in phase with the underlying signals:
in other words, if both $\ckr_k$ and $\ckr_{\ell}$ are real,
then the noise between real parts is correlated,
but the imaginary parts are uncorrelated. Thus, the figure corresponding
to\ \ref{correl} for imaginary parts would show an ellipse
centered at the origin, with principal axes aligned with the coordinate axes.

\subsection{Autocorrelation Spectra}\label{acspect_subsection}

For the autocorrelation function, the
special cases of $\tau=0$, or $\ups=0$, or both,
described by Eqs.\ \ref{ata0c} and\ \ref{a0a0c},
lead to additional correction terms that must be included in the sums.

It is useful to classify the terms in Eqs.\ \ref{ata0c} and\ \ref{a0a0c}.
Some involve a factor of the autocorrelation of the
autocorrelation function $\alp_{\tau}$,
$\sum_{n=1}^{N_o} \alp_{n+\tau} \alp_{-n}\pa$.
Others involve a simple factor of $\alp_{\tau}$,
or the product $\alp_{\tau} \alp_{-\ups}$.
Finally, some terms in the special case
$\tau=\ups=0$ do not involve $\alp$ at all: they are constants.
These 3 types of terms Fourier transform in different ways.
The additional correction terms also have terms
of the first and second sort.
The Appendix gives expressions helpful for the three sorts of Fourier transforms.

The Fourier transform of Eqs.\ \ref{A1_sum} through\ \ref{a0a0c}
yields:
\begin{eqnarray}
\la \twa_k \twa_k^*\ra &-& \la \twa_k\ra\la\twa_k^*\ra = 
{{(2 N)}\over{(2 N_o)}} \Big[2\big(A+B^2(\twa_k-1)\big)^2 \Big] \label{digital_akakc}             \label{FT_auto_1n3n7+}
\end{eqnarray}
where for simplicity I have omitted terms smaller by a factor of $1/(2 N)$ or more.
The complete expression includes additional terms of these orders,
but they are small for spectra containing more than a few channels.
Note that, again, the noise in the spectral domain can be represented by
the ``digitization noise,'' a spectrally-constant noise $(A-B)/B$ added in quadrature with 
the signal $\twa_k$.

\subsection{Correlation of Noise Across Spectral Channels}

Just as in the case of the cross-power spectrum,
the variation of noise on the correlation (explored in Paper 1) leads to correlations
in the spectral domain for the autocorrelation spectrum.
An argument precisely analogous to that for the cross-power spectrum,
in \S\ref{covar_ckrckrc} above, shows that only the second of the 3 terms in
Eq.\ \ref{A1_sum} contributes to the covariance. 
That covariance is given by:
\begin{equation}
\la \twa_k \twa_{\ell}^*\ra - \la \twa_k\ra \la\twa_{\ell}^*\ra
= {{1}\over{2 N_o}} \big[4(C-A)B^2 - 8 B^4\big] \twa_k \twa_{\ell}^* .
\label{covar_akal}
\end{equation}
This expression should be compared with Eq.\ \ref{covar_pkpl}.
The covariance is twice as great for the autocorrelation function.

\section{DISCUSSION}\label{discussion}

\subsection{Quantization Noise: One of Many Channels}

In the limit of spectra with many channels, $2N>>1$,
the noise in one particular channel is given by
terms in Eqs\ \ref{digital_ckrkckrk} and\ \ref{digital_ckrkckrkc}
with coefficient $2N$ for cross-power spectra.
In this approximation,
\begin{eqnarray}
\la \ckr_k \ckr_k^*\ra &\approx & \la \ckr_k\ra \la\ckr_k^*\ra
+{{2N}\over{2N_o}} \big[ 2 B_X^2 B_Y^2 \big]
\left(\twa_k+\left({{A_{X2}}\over{B_X^2}}-1\right)\right)
\left(\twa_k+\left({{A_{Y2}}\over{B_Y^2}}-1\right)\right)\\
\la \ckr_k \ckr_k\ra &\approx & \la \ckr_k\ra \la\ckr_k\ra
+{{2N}\over{2N_o}}
\big[ 2 B_X^2 B_Y^2 \big]\twp_k\twp_k
\end{eqnarray}
These equations closely resemble the expressions for noise for continuous
correlation, Eqs.\ \ref{continuous_spectrum_facts} and\ \ref{continuous_spectrum_facts},
except that everything has been multiplied
by the gain factor $B_X^2 B_Y^2$, and a white noise component
with variance $({{A_{X2}}\over{B_X^2}}-1)$ (or the corresponding quantity for $Y$)
has been added to the autocorrelation spectrum $\twa_k$.
These factors are those represented in the gain of the quantized cross-power spectrum (see Eq.\ \ref{avg_hat_rk}),
and in the gain and offset of the autocorrelation spectrum (see Eq.\ \ref{avg_hat_rk}).
Note that $A_{X2}>B_X^2$ for all $(v_0,\;n)$, so that the added noise component
is always positive.
This component is conveniently interpreted as quantization noise.
In this particular approximation,
treatment of the effects of quantization as white noise added in quadrature
is accurate.

\subsubsection{Correlation of Noise and Noise Reduction}

Because the noise in different spectral channels is covariant (often with negative covariance),
the integrated noise across a spectral channel is different from the summed, squared values of the noise in
each channel (often less).
Eqs.\ \ref{covar_pkplc},\ \ref{covar_pkpl}, and\ \ref{covar_akal} give the covariances.
Although the covariances
are smaller than the variances of the spectral channels given above by factors of $2 N$, they sum coherently
across the channel, whereas the variances do not.  Thus, in principle they yield comparable contributions
when summed over all channels.  
In practice, of course,
the results of such a sum are given by Eqs.\ \ref{digital_rtrtc}
and\ \ref{digital_rtrt} with $\tau=0$, 
or Eq.\ \ref{a0a0c} for autocorrelation,
because the sum over all spectral channels yields the zeroth lag.
The interested reader can verify that the results for this lag are identical to those of Paper 1, for a white spectrum.

In principle, the reduction of noise by the covariances offers the possibility of reducing quantization noise in a 
spectrally-narrow signal.  For example, one could introduce additional correlated signals, with known 
$\tilde \alpha_k$ and $\tilde \rho_k$, and measure the variation of those from theoretically-expected results.
Using Eqs.\ \ref{covar_pkplc},\ \ref{covar_pkpl}, and\ \ref{covar_akal} one can calculate what weighted sum of
those variations should be applied to the unknown signal, to reduce the noise as much as possible.  
This potential application is closely related to ``dithering'' in quantization
(see, for example, \citet{bal05} and references therein).

\subsection{Symmetries}

Note that the noise in the cross-correlation function depends on
both $\alpha$ and $\rho$.
The variance
$\la\ckr_k\ckr_k^*\ra-\la\ckr_k\ra\la\ckr_k^*\ra$
measures the summed squares of the principal axes of the elliptical Gaussian distribution
of noise, or its overall size. 
As one might expect from Eq.\ \ref{continuous_twrktwrkc}, that size depends primarily on the 
autocorrelation spectrum in that channel $\twa_k$.
The error ellipse must maintain the same size under the transformation
$\rho\rightarrow e^{i\phi}\rho$, so the noise can depend only on even powers of $\twa$, as it does.

Similarly, the variance
$\la\ckr_k\ckr_k\ra-\la\ckr_k\ra\la\ckr_k\ra$
measures the difference of the squares of the principal axes of the elliptical Gaussian distribution
of noise, or its shape. This shape must be circular for $\twp=0$, 
so that the variance must vanish there,
and so one expects that it cannot depend on $\twa$ independently of $\twp$.
Eq.\ \ref{continuous_twrktwrk} confirms this.
The difference must remain the same under the transformation $\twp\rightarrow-\twp$,
for example, so dependence on $\twp$ must be second order.

\subsection{Limits of Validity}

Numerical experiments suggest that Eqs.\ \ref{digital_ckrkckrkc} 
and\ \ref{digital_ckrkckrk} reach their limits most commonly for
spectra encountered in radio astronomy when the autocorrelation function
becomes large at lags other than the zero lag.
For example, for a single narrow line,
when the integrated power in the line becomes comparable to the
integrated continuum (including system noise),
then the autocorrelation function will reach about 0.5
in nonzero lags.
This usually leads to noise larger than that expected from the second-order analytical expressions,
especially in channels containing the line, but also throughout the spectrum.

For particular spectra, 
the additional noise can be modeled accurately by expressions that involve
higher-order terms allowed by the preceding discussion, such as $\twa_k\twp_k^2$
or $\twa_k^3$.  

\section{SUMMARY}\label{summary}

This paper investigates signal and noise for correlation of digitized data.
I assume that the received data are noiselike, in the sense that 
amplitudes and phases are drawn from complex Gaussian distributions in the spectral domain.
The variance varies with frequency.  For cross-correlation of two data streams,
covariance between the data streams may also depend on frequency.
Almost all astrophysical signals have this character.
The variances and covariances contain all the information in the signal.
The observed time series are the Fourier transforms of these spectral components.
At millimeter and longer wavelengths, these time series are commonly digitized,
and then correlated to obtain estimates of the underlying variances and covariances.
The correlation functions are finally Fourier transformed to yield the 
estimated autocorrelation or cross-power spectrum.
Averaged over a number of realizations, the elements of the correlation function will approach
a Gaussian distribution.
The mean correlation
represents the deterministic part of a measurement, or 
the signal.  The standard deviation of the measurement represents the random part, or noise.

Digitization of the signals involves quantization, which represents the 
continuous signal with a finite set of levels, and thus destroys information.
This affects both the signal and the noise.
I summarize results for continuous data in \S\ref{continuous},
and present new results, for noise for quantized data, in \S\ref{quantized} 
and \S\ref{section_quantspectrum}.

In \S\ref{quantized} I investigate statistics of correlation functions.
Under the assumption that the correlation is smaller than 1
(except equal to 1 for the zero-lag of the autocorrelation function), I find expressions for the mean
cross- and autocorrelation functions.
Results agree with earlier work \citep{coo70,jen98}.
I then find analytical expressions for the noise in the correlation functions.
This noise takes the form of variances of the measured elements, as a function of lag;
and of covariances between the measured elements.

In \S\ref{section_quantspectrum} I investigate statistics of 
spectra.
The mean spectra are related to the mean correlation functions by Fourier transform;
the noise in the spectra is related to that in the correlation functions by a double Fourier transform.
I find that 
the mean cross-power spectrum for quantized data equals that for
continuous data, times a gain factor. The mean autocorrelation spectrum 
equals that for continuous data times the same gain factor, plus white noise added in
quadrature with the original data:
``quantization noise''.
This accords with previous results \citep{coo70,jen98}.
I then find analytical expressions for the noise in the spectra.
For both cross-power and autocorrelation spectra,
I find that noise in one channel of a spectrum 
is equal to a gain factor times that for continuous data, plus
the same quantization noise found for the autocorrelation spectrum.
However, I also find that noise is correlated (most commonly anticorrelated) across spectral channels.
Thus, when noise increases the value measured in one channel above the mean,
noise will tend to decrease the value measured in another channel.
This correlation can produce a contribution comparable to, or even greater than,
the quantization noise when summed over all spectral channels.

\acknowledgments

I am grateful 
to the DRAO for supporting this work with extensive
correlator time.
I gratefully acknowledge the VSOP Project, which is led by the Japanese 
Institute of Space and Astronautical Science in cooperation with many 
organizations and radio telescopes around the world.
The U.S. National Science Foundation provided partial financial support for this work.

\appendix

\section{Useful Facts for Spectra}\label{useful_spectral_facts}

Parseval's theorem states:
\begin{equation}
\sum_{\tau=-N}^{N-1} \rho_{\tau}\rho_{\tau}^* 
= {{1}\over{2N}} \sum_{k=-N}^{N-1}\twp_k\twp_k^* .
\label{parseval}
\end{equation}

Therefore,
\begin{equation}
\sum_{\tau\neq\ups}
e^{i{{2\pi}\over{2N}}k(\tau-\ups)}\rho_{\tau}\rho_{\ups}^*=
\twp_k \twp_k^*
-
{{1}\over{(2N)}}\sum_{k=-N}^{N-1}\twp_k\twp_k^* .
\end{equation}
Also note that
\begin{equation}
\sum_{\tau\neq\ups}
e^{i{{2\pi}\over{2N}}k(\tau-\ups)} \alp_{(\tau-\ups)} 
= 
(2N) \twa_k - (2N)
\end{equation}

\begin{equation}
\sum_{\nu=-N}^{N-1} e^{i{{2\pi}\over{2 N}}\nu k} \sum_{n=1}^{N_o}\alp_{n}\alp_{-n+\nu}
=(2N) \twa_k^2.
\label{convolution_fact}
\end{equation}

For convolutions, recall that
\begin{equation}
\sum_{\tau=-N}^{N-1}\sum_{\ups=-N}^{N-1}
e^{i{{2\pi}\over{2N}}k(\tau-\ups)}
\sum_{n=1}^{N_o} \rho_{n}\rho_{-n+(\tau-\ups)}
= (2N) \twp_k \twp_k .
\label{convolution_fact_rho}
\end{equation}
where I assume that the correlation function wraps,
and that the correlation function includes all lags
with nonzero signal.

I define the quantity:
\begin{equation}
\tilde C_{\rho}(k)=\sum_{\tau=-N}^{N-1} e^{i{{2\pi}\over{2N}} 2 k \tau} \rho_{\tau} \rho_{\tau} .
\end{equation}

\newpage
\figurenum{1}

\newpage
\figurenum{2}

\newpage
\figurenum{3}

\newpage
\figurenum{4}

\newpage
\figurenum{5}

\newpage
\figurenum{6}

\clearpage

\begin{deluxetable}{clll}
\tablenum{1}
\tablecolumns{4}
\tablewidth{0pc}
\tablecaption{Second and Fourth Moments of Quantized Real Gaussian Variables}
%
%
\tablehead{
& \colhead{Terms of 2nd or Lower Order} & \colhead{Moments of} \\ 
\colhead{Average}  &\colhead{that Contribute to Expansion}& \colhead{Quantized Variables} &  \\ 
}
\startdata
$\la \hat W \hat X               \ra    $&$ [W X] (\rWX)                           $&$ [B_{W} B_{X}]                \rWX                               $\\
$\la \hat W^2                    \ra    $&$ [1]                                    $&$ [A_{W2}]                                                        $\\
$                                       $&$                                        $&$                                                                 $\\
$\la \hat W \hat X \hat Y \hat Z \ra    $&$ [W X Y Z] (\rWX\rYZ+\rWY\rXZ+\rWZ\rXY) $&$ [B_{W} B_{X} B_{Y} B_{Z}] (\rWX\rYZ+\rWY\rXZ+\rWZ\rXY)          $\\
$\la \hat W^2 \hat X \hat Y      \ra    $&$ [X Y] (\rXY)+ [(W^2-1) X Y] (\rWX\rWY) $&$ [A_{W2}B_{X} B_{Y}] (\rXY)+[(C_{W2}-A_{W2})B_{X}B_{Y}](\rWX\rWY)$\\
$\la \hat W^2 \hat X^2           \ra    $&$ [1] + [\haf (1-W^2)(1-X^2)] (\rWX^2)   $&$ [A_{W2}A_{X2}]+[\haf(C_{W2}-A_{W2})(C_{X2}-A_{X2})]\rWX^2       $\\
$\la \hat W^3 \hat X             \ra^{a}$&$ [W X] (\rWX)                           $&$ [B_{W3}B_{X}]                \rWX                               $\\
$\la \hat W^4                    \ra^{a}$&$ [1]                                    $&$ [A_{W4}]                                                        $\\
\enddata
\tablenotetext{a} {Important only for autocorrelations.}
\label{table_real_avgs}
\end{deluxetable}
\clearpage

\begin{deluxetable}{lll}
\tablenum{2}
\tablecolumns{3}
\tablewidth{0pc}
\tablecaption{Fourth Moments of Quantized Complex Gaussian Variables:
$\hat w$ $\hat x$ $\hat y$ and $\hat z$ }
\tablehead{
\colhead{Class} & \colhead{Form} & \colhead{Result: Quantized} \\ 
}
\startdata
{\scriptsize $11$}   &$\la \hwh \hxc \ra               $&$ [2 B_W B_X] \rwx $\\
                     &$                                $&$                  $\\
{\scriptsize $2$}    &$\la \hwh \hwc \ra ^{a}          $&$ [2 A_{W2}]       $\\
                     &$                                $&$                  $\\
{\scriptsize $1111+$}&$\la \hwh \hxc \hyc \hzh \ra     $&$ [4 B_W B_X B_Y B_Z] (\rwx\ryz^*+\rwy\rxz^*) $\\
{\scriptsize $1111-$}&$\la \hwh \hxc \hyh \hzc \ra     $&$ [4 B_W B_X B_Y B_Z] (\rwx\ryz  +\rwz\rxy^*) $\\
                     &$                                $&$                  $\\
{\scriptsize $211+$} &$\la \hwh \hxc \hwc \hyh  \ra    $&$ [2 (C_{W2}-A_{W2})B_X B_Y](\rwx\rwy^*)+[4 A_{W2}B_X B_Y](\rxy^*)$\\
{\scriptsize $211-$} &$\la \hwh \hxc \hwh \hyc  \ra    $&$ [2 (C_{W2}-A_{W2})B_X B_Y + 4 B_W^2 B_X B_Y ](\rwx\rwy)$\\
                                              
                     &$                                $&$                    $\\
{\scriptsize $22+$}  &$\la \hwh \hxc \hwc \hxh  \ra    $&$ [(C_{W2}-A_{W2})(C_{X2}-A_{X2})](\rwx \rwx^*) + [4 A_{W2} A_{X2}] $\\
{\scriptsize $22-$}  &$\la \hwh \hxc \hwh \hxc  \ra    $&$\phantom{+}[\haf ((C_{W2}-A_{W2})+2 B_W^2)((C_{X2}-A_{X2})+2 B_X^2)] (\rwx   \rwx  )    $\\
                     &$                                $&$         + [\haf ((C_{W2}-A_{W2})-2 B_W^2)((C_{X2}-A_{X2})-2 B_X^2)] (\rwx^* \rwx^*)    $\\
                                              
                     &$                                          $&$                    $\\
{\scriptsize $31+$}  &$\la \hwh \hwc \hwc \hxh  \ra^{a}$&$ [ 2 B_{W3} B_X + 2 A_{W2} B_W B_X ] \rwx^* $\\
{\scriptsize $31-$}  &$\la \hwh \hwc \hwh \hxc  \ra^{a}$&$ [ 2 B_{W3} B_X + 2 A_{W2} B_W B_X ] \rwx   $\\
                                             
                     &$                                $&$                          $\\
{\scriptsize $4$}    &$\la \hwh \hwc \hwc \hwh  \ra^{a}$&$ [ 2 A_{W4} + 2 A_{W2}^2] $\\ 
                     &$                                $&$                          $\\
\enddata
\tablenotetext{a} {Important only for autocorrelations.}
\label{table_complex_avgs}
\end{deluxetable}
\clearpage

\begin{deluxetable}{llllllllcl}
\tablenum{3}
\tablecolumns{10}
\tablewidth{0pc}
\tablecaption{Terms in XCF Sums:  $\hat r\hat r^*$ and $\hat r\hat r$}
\tablehead{
\colhead{ID}         &
\colhead{Conditions} & \multicolumn{4}{c}{Subscript} &
\colhead{Class}      & \colhead{Form}  & \colhead{Multiplicity} & \colhead{Notes} \\
\cline{3-6}\\
\colhead{}           &
\colhead{}           & \colhead{$a$}   & \colhead{$b$}          & \colhead{$c$}   & \colhead{$d$} &
\colhead{}           & \colhead{}      & \colhead{}             & \colhead{}
}
\startdata
\cutinhead{Xc: $\la\hrh_{\tau}\hrc_{\ups} \ra=\la\hxh_{\ell} \hyc_{\ell+\tau} \hxc_{m} \hyh_{m+\ups}\ra=\la\hxh_a\hyc_b\hxc_c\hyh_d\ra$ }
\sidehead{Xcn: $\tau\neq\ups$:}
Xcn.1 &$ \ell=m            $&$ \ell $&$ \ell+\tau $&$\ell            $&$\ell+\ups $&\scriptsize{$211+ $}&$\la \hwh \hxc \hwc \hyh \ra$&$ N_o        $& Eq.\ 41   \\
Xcn.2 &$ \ell+\tau= m+\ups $&$ \ell $&$ \ell+\tau $&$\ell+(\tau-\ups)$&$\ell+\tau $&\scriptsize{$211+ $}&$\la \hwh \hxc \hwc \hyh \ra$&$ N_o        $&Eq.\ 42,a \\
Xcn.3 &$ \ell+\tau= m      $&        &             &                  &            &\scriptsize{$1111+$}&$\la \hwh \hxc \hyc \hzh \ra$&$ -          $&  b   \\
Xcn.4 &$ \ell= m+\ups      $&        &             &                  &            &\scriptsize{$1111+$}&$\la \hwh \hxc \hyc \hzh \ra$&$ -          $&  b   \\
Xcn.0 &4 distinct           &$ \ell $&$ \ell+\tau $&$ m              $&$ m  +\ups $&\scriptsize{$1111+$}&$\la \hwh \hxc \hyc \hzh \ra$&$ N_o^2-2N_o $&Eq.\ 40   \\

\sidehead{Xce: $\tau =\ups $}
Xce.1 &$ \ell=m            $&$ \ell $&$ \ell+\tau $&$ \ell           $&$\ell+\tau $&\scriptsize{$22+  $}&$\la \hwh \hxc \hwc \hxh \ra$&$ N_o        $&   \\
Xce.0 &4 distinct           &$ \ell $&$ \ell+\tau $&$ m              $&$ m  +\tau $&\scriptsize{$1111+$}&$\la \hwh \hxc \hyc \hzh \ra$&$ N_o^2-N_o  $&   \\
                                                                                                                                     
\cutinhead{Xr: $\la\hrh_{\tau}\hrh_{\ups}\ra=\la\hxh_{\ell}\hyc_{\ell+\tau}\hxh_{m}\hyc_{m+\ups}\ra=\la\hxh_a\hyc_b\hxh_c\hyc_d\ra$}
\sidehead{Xrn: $\tau\neq\ups$:}                                                                                                      
Xrn.1 &$ \ell=m            $&$ \ell $&$ \ell+\tau $&$\ell            $&$\ell+\ups $&\scriptsize{$211- $}&$\la \hwh \hxc \hwh \hyc \ra$&$ N_o        $&   \\
Xrn.2 &$ \ell+\tau= m+\ups $&$ \ell $&$ \ell+\tau $&$\ell+(\tau-\ups)$&$\ell+\tau $&\scriptsize{$211- $}&$\la \hwh \hxc \hyh \hwc \ra$&$ N_o        $&  a\\
Xrn.3 &$ \ell+\tau= m      $&        &             &                  &            &\scriptsize{$1111-$}&$\la \hwh \hxc \hyh \hzc \ra$&$ -          $&  b\\
Xrn.4 &$ \ell= m+\ups      $&        &             &                  &            &\scriptsize{$1111-$}&$\la \hwh \hxc \hyh \hzc \ra$&$ -          $&  b\\
Xrn.0 &4 distinct           &$ \ell $&$ \ell+\tau $&$ m              $&$ m  +\ups $&\scriptsize{$1111-$}&$\la \hwh \hxc \hyh \hzc \ra$&$ N_o^2-2N_o $&   \\
                                                                                                                                     
\sidehead{Xre: $\tau =\ups$}                                                                                                         
Xre.1 &$ \ell=m            $&$ \ell $&$ \ell+\tau $&$ \ell           $&$\ell+\tau $&\scriptsize{$22-  $}&$\la \hwh \hxc \hwh \hxc \ra$&$ N_o        $&   \\
Xre.0 &4 distinct           &$ \ell $&$ \ell+\tau $&$ m              $&$ m  +\tau $&\scriptsize{$1111-$}&$\la \hwh \hxc \hyh \hzc \ra$&$ N_o^2-N_o  $&   \\

\enddata
\tablenotetext{a} {Roles of $\hat x$ and $\hat y$ reversed from Table\ \ref{table_complex_avgs}.   
Use complex conjugate. }
\tablenotetext{b} {Important only for autocorrelations. Yields standard form for cross-correlation.}
\label{table_terms_in_Xsums}
\end{deluxetable}
\clearpage

\begin{deluxetable}{llllllllcl}
\tablenum{4}
\tablecolumns{10}
\tablewidth{0pc}
\tablecaption{Terms in ACF Sums: $\hat a\hat a^*$ }
\tablehead{
\colhead{ID}         &
\colhead{Conditions} & \multicolumn{4}{c}{Subscript} &
\colhead{Class}      & \colhead{Form}  & \colhead{Multiplicity} & \colhead{Notes} \\
\cline{3-6}\\
\colhead{}           &
\colhead{}           & \colhead{$a$}   & \colhead{$b$}          & \colhead{$c$}   & \colhead{$d$} &
\colhead{}           & \colhead{}      & \colhead{}             & \colhead{}
}
\startdata
\cutinhead{ACF: $\la\hah_{\tau}\hac_{\ups}\ra=\la\hxh_{\ell} \hxc_{\ell+\tau} \hxc_{m} \hxh_{m+\ups}\ra=\la\hxh_a\hxc_b\hxc_c\hxh_d\ra$}
\sidehead{Antu: $\tau\neq\ups$, $\tau\neq 0$, $\ups\neq 0$}
Antu.1 &$ \ell=m            $&$ \ell $&$ \ell+\tau $&$\ell            $&$\ell+\ups     $&\scriptsize{$ 211+   $}&$\la \hwh \hxc \hwc \hyh \ra$&$ N_o        $&  \\
Antu.2 &$ \ell+\tau= m+\ups $&$ \ell $&$ \ell+\tau $&$\ell+(\tau-\ups)$&$\ell+\tau     $&\scriptsize{$(211+)^*$}&$\la \hxh \hwc \hyc \hwh \ra$&$ N_o        $& a\\
Antu.3 &$ \ell+\tau= m      $&$ \ell $&$ \ell+\tau $&$ \ell+\tau      $&$\ell+\tau+\ups$&\scriptsize{$(211-)^*$}&$\la \hxh \hwc \hwc \hyh \ra$&$ N_o        $& a\\ 
Antu.4 &$ \ell= m+\ups      $&$ \ell $&$ \ell+\tau $&$ \ell-\ups      $&$\ell          $&\scriptsize{$ 211-   $}&$\la \hwh \hxc \hyc \hwh \ra$&$ N_o        $&  \\
Antu.0 &4 distinct           &$ \ell $&$ \ell+\tau $&$ m              $&$ m  +\ups     $&\scriptsize{$ 1111+  $}&$\la \hwh \hxc \hyc \hzh \ra$&$ N_o^2-4N_o $&  \\
                                                                                                                                                
\sidehead{An0u: $\tau\neq\ups$, $\tau=0$:}                                                                                                      
An0u.1&$ \ell=m             $&$ \ell $&$ \ell      $&$ \ell           $&$\ell+\ups     $&\scriptsize{$ 31+    $}&$\la \hwh \hwc \hwc \hxh \ra$&$ N_o        $   \\
An0u.2&$ \ell=m+\ups        $&$ \ell $&$ \ell      $&$ \ell-\ups      $&$ \ell         $&\scriptsize{$ 31-    $}&$\la \hwh \hwc \hxc \hwh \ra$&$ N_o        $   \\
An0u.0&3 distinct            &$ \ell $&$ \ell      $&$ m              $&$ m  +\ups     $&\scriptsize{$ 211+   $}&$\la \hwh \hwc \hxc \hyh \ra$&$ N_o^2-2N_o  $  \\
											         			      
\sidehead{Ant0: $\tau\neq\ups$, $\ups=0$:}						         			      
Ant0.1&$ \ell=m             $&$ \ell $&$ \ell+\tau $&$ \ell           $&$\ell          $&\scriptsize{$ 31-    $}&$\la \hwh \hxc \hwc \hwh \ra$&$ N_o        $&  \\
Ant0.2&$ \ell+\tau= m       $&$ \ell $&$ \ell+\tau $&$ \ell+\tau      $&$ \ell+\tau    $&\scriptsize{$ 31+    $}&$\la \hxh \hwc \hwc \hwh \ra$&$ N_o        $&  \\
Ant0.0&3 distinct            &$ \ell $&$ \ell+\tau $&$ m              $&$ m            $&\scriptsize{$ 211+   $}&$\la \hxh \hyc \hwc \hwh \ra$&$ N_o^2-2N_o $&  \\
											         			      
\sidehead{Aet: $\tau =\ups$, $\tau\neq 0$:}						         			      
Aet.1 &$ \ell=m             $&$ \ell $&$ \ell+\tau $&$ \ell           $&$\ell+\tau     $&\scriptsize{$ 22+    $}&$\la \hwh \hxc \hwc \hxh \ra$&$ N_o        $&  \\
Aet.2 &$ \ell=m+\tau        $&$ \ell $&$ \ell+\tau $&$ \ell-\tau      $&$\ell          $&\scriptsize{$ 211-   $}&$\la \hwh \hxc \hyc \hwh \ra$&$ N_o        $& a\\
Aet.3 &$ \ell+\tau=m        $&$ \ell $&$ \ell+\tau $&$ \ell+\tau      $&$\ell+2\tau    $&\scriptsize{$(211-)^*$}&$\la \hxh \hwc \hwc \hyh \ra$&$ N_o        $& a\\
Aet.0 &4 distinct            &$ \ell $&$ \ell+\tau $&$ m              $&$ m  +\tau     $&\scriptsize{$ 1111+  $}&$\la \hwh \hxc \hyc \hzh \ra$&$ N_o^2-3N_o $&  \\
											         			      
\sidehead{Ae0: $\tau =\ups$, $\tau=0$:}							         			      
Ae0.1&$ \ell=m              $&$ \ell $&$ \ell      $&$ \ell           $&$\ell          $&\scriptsize{$ 4      $}&$\la \hwh \hwc \hwc \hwh \ra$&$ N_o        $&  \\
Ae0.0&2 distinct             &$ \ell $&$ \ell      $&$ m              $&$ m            $&\scriptsize{$ 22+    $}&$\la \hwh \hwc \hxc \hxh \ra$&$ N_o^2-N_o  $&  \\
\enddata
\tablenotetext{a} {Complex conjugate of element in Table\ \ref{table_complex_avgs}.}
\label{table_terms_in_Asums}
\end{deluxetable}
\clearpage

\clearpage

\begin{thebibliography}{99}
\bibitem[Balestrieri et al.(2005)]{bal05}Balestrieri, E., Daponte, P., \& Rapuano, S. 2005, IEEE Trans. Inst. Meas., 54, 1388
\bibitem[Cooper(1970)]{coo70}Cooper, B.F.C. 1970, AustJPhys, 23, 521 
\bibitem[D'Addario et al.(1984)]{dad84}D'Addario, L.R., Thompson, A.R., Schwab, F.R., \& Granlund, J. 1984, Radio Sci., 19, 931
\bibitem[Dennett-Thorpe \& de Bruyn(2002)]{den02}Dennett-Thorpe, J., \& de Bruyn, A. G. 2002, Nature, 415, 57.
\bibitem[Desai et al.(1992)]{des92}Desai, K.M., Gwinn, C.R., Reynolds, J.R., King, E.A., Jauncey, D., Flanagan, C.,  Nicolson, G., Preston, R.A., \& Jones, D.L. 1992, ApJ, 393, L75
\bibitem[Gwinn et al.(2000)]{gwi00}Gwinn, C.R., Britton, M.C., Reynolds, J.E., Jauncey, D.L., King, E.A., McCulloch, P.M., Lovell, J.E.J., Flanagan, C.S., Preston, R.A. 2000, ApJ, 531, 902 
\bibitem[Gwinn(2004)]{gwi04}Gwinn, C.R. 2004, PASP, 116, 84 (Paper 1)
\bibitem[Hagen \& Farley(1973)]{hag73}Hagen, J.B., \& Farley, D.T. 1973, Radio Sci, 8, 775 
\bibitem[Jauncey et al.(2000)]{jau00}Jauncey, D.L., Kedziora-Chudczer, L.L., Lovell, J.E.J., Nicolson, G.D., Perley, R.A., Reynolds, J.E., Tzoumis, A.K., Wieringa, M.H. 2000, Astrophysical Phenomena Revealed by Space VLBI, eds. H. Hirabayashi, P.G. Edwards, D.W. Murphy, ISAS: Sagamihara, p. 147
\bibitem[Jenet \& Anderson(1998)]{jen98}Jenet, F.A., \& Anderson, S.B. 1998, PASP, 110, 1467 
\bibitem[Kulkarni \& Heiles(1980)]{kul80}Kulkarni, S.R., \& Heiles, C. 1980, AJ, 85, 1413 
\end{thebibliography}
\end{document}